\documentclass[superscriptaddress,aps,journal=prl,preprint]{revtex4-2}
\usepackage{graphicx}
\usepackage{epstopdf}
\usepackage{bm,amsmath,amssymb} 
\usepackage{color} 
\usepackage{dcolumn}   
\usepackage{multirow}
\usepackage{hyperref}
\newcommand{\etal}{{\em et al.\ }}

\newcommand{\nL}{{$n_{\rm L}$}}

\begin{document}
\title{Depolarization Induced III-V Triatomic Layers with Tristable  Polarization States}%

\author{Changming Ke}
\affiliation{Key Laboratory for Quantum Materials of Zhejiang Province, Department of Physics, School of Science, Westlake University, Hangzhou, Zhejiang 310030, China}
\affiliation{Institute of Natural Sciences, Westlake Institute for Advanced Study, Hangzhou, Zhejiang 310024, China}
\author{Yihao Hu}
\affiliation{Key Laboratory for Quantum Materials of Zhejiang Province, Department of Physics, School of Science, Westlake University, Hangzhou, Zhejiang 310030, China}
\affiliation{Institute of Natural Sciences, Westlake Institute for Advanced Study, Hangzhou, Zhejiang 310024, China}
\author{Shi Liu}
\email{liushi@westlake.edu.cn}
\affiliation{Key Laboratory for Quantum Materials of Zhejiang Province, Department of Physics, School of Science, Westlake University, Hangzhou, Zhejiang 310030, China}
\affiliation{Institute of Natural Sciences, Westlake Institute for Advanced Study, Hangzhou, Zhejiang 310024, China}


\begin{abstract} 
The integration of ferroelectrics that exhibit high dielectric, piezoelectric, and thermal susceptibilities with the mainstream semiconductor industry will enable novel device types for widespread applications, and yet there are few silicon-compatible ferroelectrics suitable for device downscaling. We demonstrate with first-principles calculations that the enhanced depolarization field at the nanoscale can be utilized to soften unswitchable wurtzite III-V semiconductors, resulting in ultrathin two-dimensional (2D) sheets possessing reversible polarization states. A 2D sheet of AlSb consisting of three atomic planes is identified to host both ferroelectricity and antiferroelectricity, and the tristate switching is accompanied by a metal-semiconductor transition. 
The thermodynamics stability and potential synthesizability of the triatomic layer are corroborated with phonon spectrum calculations, {\em ab initio} molecular dynamics, and variable-composition evolutionary structure search.  
We propose a 2D AlSb-based homojunction field effect transistor that supports three distinct and nonvolatile resistance states. This new class of III-V semiconductor-derived 2D materials with dual ferroelectricity and antiferroelectricity opens up the possibility for nonvolatile multibit-based integrated nanoelectronics.

\end{abstract}

\maketitle
\newpage
Ferroelectricity, as an extensively studied dipolar ordering state of insulators, is characterized by electrically switchable polarization. The strong coupling between polarization, strain, and electronic degrees of freedom of ferroelectrics have made them critical components in numerous devices such as sensors, actuators, and nonvolatile memories~\cite{Scott07p954,Martin16p16087}. The continuing demand for miniaturized electronics has imposed stringent requirements on ferroelectrics. 
In particular, to incorporate ferroelectric functionalities into integrated circuits via the current semiconductor manufacturing process, materials with nanoscale switchable dipoles and silicon compatibility are essential~\cite{Mikolajick21p100901}. 
 
Two-dimensional (2D) ferroelectrics with long-range dipolar ordering in atom-thick crystalline layers are promising materials for ferroelectric-based nanoelectronics because of their various merits such as the uniform atomic thickness for high-density integration and the easy preparation of high-quality interface in van der Waals heterostructures~\cite{Wu18pe1365}. However, similar to perovskite ferroelectrics, most 2D ferroelectrics also suffer from the depolarization effect such that they often have the polarization developed in-plane~\cite{Guan19p1900818, Kruse22arxiv}, a feature that is inconvenient for lateral downscaling. Atomically thin monolayers with out-of-plane polarization ($P_{\rm OP}$) 
remains rare, and few notable examples confirmed experimentally are CuInP$_2$S$_6$~\cite{Liu16p12357}, $\alpha$-In$_2$Se$_3$~\cite{Ding17p14956,Zhou17p5508,Cui18p1253,Xiao18p227601,Poh18p6340,Wan18p14885}, MoTe$_2$~\cite{Yuan19p1775}, and WTe$_2$~\cite{Fei18p336}. Additionally, it remains unclear how to integrate these 2D ferroelectrics with the mainstream semiconductor technology. 

A strategy to obtain new ferroelectrics suitable for integrated systems is to ``soften" silicon-compatible piezoelectrics to make them switchable by applying appropriate ``stressors"~\cite{Ferri21p044101}. For example, by substituting Sc into a well-known nitride piezoelectric, AlN, Fichtner \etal~realized a giant switchable polarization (80--110 $\mu$C/cm$^2$) in Al$_{1-x}$Sc$_x$N~\cite{Fichtner19p114103}. More recently, starting with another widely used piezoelectric, ZnO, Ferri~\etal synthesized thin films of Zn$_{1-x}$Mg$_{x}$O  and reported even larger switchable polarization of $>100$~$\mu$C/cm$^{2}$ and coercive fields below 3~MV/cm at room temperatures~\cite{Ferri21p044101}. In both cases, the essence is to destabilize an unswitchble piezoelectric by applying a chemical stressor.

We propose to ``physically soften" silicon-compatible piezoelectrics represented by III-V wurtzite piezoelectrics via dimension reduction. Products based on III-V semiconductors have been widely employed in mobile devices, wireless networks, satellite communications, and optoelectronics~\cite{jeong22p9031, chen16p307, Vurgaftman01p5815}. For example, the 4th-generation (4G) wireless networks depend on thin-film bulk acoustic resonators consisting of piezoelectric wurtzite AlN. At present, the industry of III-V semiconductor manufacturing is well established. Several approaches such as direct growth of III-V on Si, III-V on lattice engineered substrate, and III-V on Ge-Si template have been developed to integrate III-V compounds with the cutting-edge modern complementary metal oxide semiconductor (CMOS) technology~\cite{yang19p101305, IIIV-on-Si}. Therefore, III-V semiconductor-based 2D ferroelectrics, if available, will reduce the barrier of integrating ferroelectric functionalities with silicon-based technology and lower the cost of commercialization.

\begin{figure}[htp]
    \centering
    \includegraphics[scale=0.45]{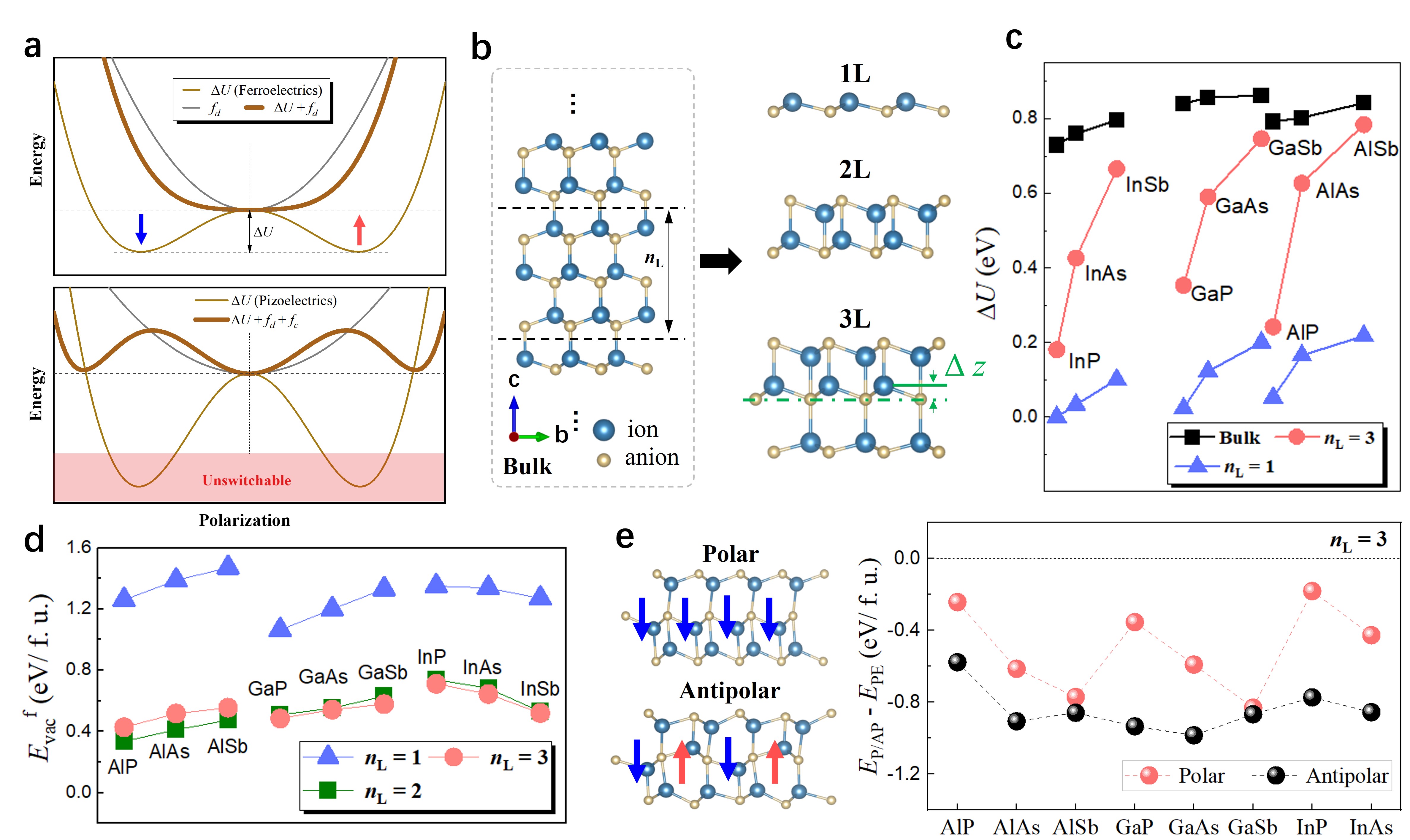}
    \caption{(a) Utilizing the depolarization energy ($f_d$) to soften unswitchable piezoelectrics with large barrier $\Delta U$ separating two polar states. The delicate balance between $\Delta U$, $f_d$, and the energy cost ($f_c$) to form antiparallel neighboring dipoles may lead to a triple well in thin films. (b) Construction of ultrathin 2D sheets by cutting wurtzite III-V piezoelectrics along the $c$ plane. The thickness of the sheet is defined as the number (\nL) of atomic planes. The right panel shows the optimized structures of monolayer (1L), diatomic layer (2L), and triatomic layer (3L). 
    (c) Energy barrier ($\Delta U$) in bulk and 1L and 3L sheets. (d) Formation energy ($E_{\rm vac}^f$) of 1L, 2L, and 3L sheets. (e) Energy of antipolar ($E_{\rm AP}$) and polar ($E_{\rm P}$) phases in 3L sheets relative to the paraelectric ($E_{\rm PE}$) phase. 
    }
    \label{design}
\end{figure}

The physical stressor we employ is the enhanced depolarization field at the nanosale. 
The depolarization field ($E_d$) arising from the incomplete screening of surface polarization bound charges scales inversely with the film thickness ($E_d\propto P_s/d$ with $P_s$ the remnant polarization and $d$ the film thickness)~\cite{Junquera03p506}. In thin films of conventional perovskite ferroelectrics such as PbTiO$_3$, the intrinsic double-well energy landscape of a ferroelectric will eventually be flattened out by the pronounced depolarization field in thin films below a critical thickness, leading to a nonpolar paraelectric ground state (Fig.~\ref{design}a top panel). In contrast, some piezoelectrics such as wurtzite AlN are unswitchable in bulk because the barrier ($\Delta U$) separating two polar states is prohibitively large such that the switching field exceeds the dielectric breakdown limit. 
Utilizing the increased depolarization energy ($f_d$) with reduced dimension ($f_d\propto P_s^2/d$) to compensate $\Delta U$, we suggest it is feasible to soften piezoelectrics to 2D ferroelectrics with switchable $P_{\rm OP}$. Another competing phase that could emerge in thin films is an antipolar phase with neighboring antiparallel dipoles that has zero depolarization energy. It is expected that the neighboring dipoles in bulk wurtzite piezoelectrics strongly favor the parallel alignment with a coupling strength characterized by $J$; forming an antipolar phase thus comes with an energy cost, $f_c\propto zJ$, with $z$ the coordination number of a local dipole. Heuristically, the competition between $\Delta U$, $f_d$, and $f_c$ determines the ground state (polar, antipolar, or paraelectric) in free-standing thin films. Moreover, a triple potential well may emerge by engineering the relative magnitudes of competing energy terms (Fig.~\ref{design}a bottom panel). 

We explore our design principle with first-principles density functional theory (DFT) calculations, focusing on ultrathin 2D sheets of wurtzite III-V compounds (III=Al, Ga, In; V=N, P, As). 
We discover a nonpolar diatomic layer (2L) and a triatomic layer (3L) with spontaneous local inversion symmetry breaking. Specifically for 3L sheets, it can adopt a high-energy polar state with $P_{\rm OP}$ and a low-energy antipolar state with neighboring antiparallel dipoles. Interestingly, the polar and antipolar states in 3L AlSb are both dynamically stable, as confirmed by phonon spectrum calculations and {\em ab initio} molecular dynamics (AIMD), and these two states are comparable in energy, making 3L AlSb an unusual tristable system that supports both ferroelectricity and antiferroelectricity. Moreover, the electronic degree of freedom is directly coupled to the polar ordering in 3L AlSb, and the tristate switching is accompanied with a metal-semiconductor transition. We propose a 2D homojunction field effect transistor (FET) consisting of 2L and 3L AlSb. The carrier type and density in the semiconducting channel of 2L AlSb can be effectively regulated by the polarization state of 3L AlSb, leading to three distinct and nonvolatile resistance states. The deterministic ferroelectric domain engineering at the nanoscale could be used to pattern the 2L-3L homojunction as high-density periodic arrays of $p$-$n$ junctions and $p$-$i$-$n$ junctions. The proposed 2D sheets of III-V compounds supporting tristable polarization states offer promise for experimental investigation and for the development and design of nonvolatile multistate functional applications such as high-density memory and synaptic electronics. 

DFT calculations are performed using Vienna {\em ab initio} Simulation Package (\texttt{VASP})~\cite{Kresse96p11169,Kresse96p15}. The interaction between the core ion and electrons is described by the projector augmented wave (PAW) method~\cite{Blochi94p17953}. The PBEsol functional is chosen as the exchange-correlation functional~\cite{Perdew08p136406}. The vacuum layer along the $c$ axis is thicker than 15~\AA~in the slab model, and the dipole correction is employed to remove the spurious interaction between different periodic images. We use an energy cutoff of 700~eV, a 8$\times$8$\times$1 Monkhorst–Pack $k$-point mesh, and an energy convergence threshold of 10$^{-8}$ eV for electronic self-consistent calculations. The convergence criterion for an optimized structure is 10$^{-7}$~eV in energy. The structural stability at finite temperatures is studied by $NVT$ AIMD simulations using the $\Gamma$-point sampling with the temperature controlled using the Nos\'e-Hoover thermostat~\cite{Nose84p511, Hoover85p1695}. The phonon spectrum is computed using the frozen phonon approach as implemented in \texttt{Phonopy}~\cite{Togo15p1} in conjunction with \texttt{VASP}.

The 2D sheet is constructed by cutting the bulk along the $c$ plane, and the thickness of the film is defined as the number (\nL) of atomic planes (Fig.~\ref{design}b). In the case of monolayers (\nL=1), we find that all nitrides favor the planar structure~\cite{Wu11p236101} whereas monolayers of other III-V compounds are buckled honeycomb structures characterized by the presence of $P_{\rm OP}$ and small values of $\Delta U$ ($<0.2$~eV, Fig.~\ref{design}c). We note that III–V buckled honeycomb monolayers have been studied previously  with DFT~\cite{Sahin09p155453, Zhuang13p165415, Gao17p1600412}, though the 2D ferroelectricity was not appreciated. The formation energy per formula unit (f.u.) of an isolated 2D sheet with respect to the bulk counterpart is defined as $E_{\rm vac}^{\rm f}=E_{\rm 2D}/n_{\rm L}-E_{\rm 3D}/N_{\rm 3D}$, where $E_{\rm 2D}$ is the energy of the 2D sheet consisting of \nL~atomic planes and $E_{\rm 3D}$ is the energy of a cell in bulk containing $N_{\rm 3D}$~atomic planes.  Though several III-V monolayers are potential 2D ferroelectrics featuring small switching barriers, their formation energies are rather large ($>0.8$~eV/f.u., Fig.~\ref{design}d), hinting at the difficulty of synthesis in experiments. 

When the thickness increases to \nL=2, for III-V compounds (III=Al, Ga, In, V=P, As, Sb), the initial wurtzite-like configuration is no longer stable, and the optimized diatomic layer denoted as 2L acquires the inversion symmetry thus being nonpolar (Fig.~\ref{design}b). 
Further increasing the thickness to \nL=3~surprisingly revives $P_{\rm OP}$. The triatomic layer labeled as 3L has a buckled  central layer that breaks the out-of-plane inversion symmetry (Fig.~\ref{design}b). Structurally, both 2L and 3L have group-V anions being the outermost surface layers. We note that 3L sheets have much lower formation energies than monolayers albeit with higher $\Delta U$ (Fig.~\ref{design}c-d).  This seems to suggest III-V compounds in the 3L form are easier to synthesize but remain unswitchable. 

Following the design principle, we investigate possible competing antipolar phases in 3L sheets. 
We identify an antipolar phase that has the energy consistently lower than the polar phase (Fig.~\ref{design}e and Fig.~S1 in Supporting Information).
Based on Shirane's energetic criterion on antiferroelectricity, an antiferroelectric is an antipolar crystal with free energy comparable to that of the reference polar crystal that has aligned sublattice local dipoles~\cite{Shirane52p219}. Therefore, we suggest 3L sheets of AlSb and GaSb likely host antiferroelectricity as their polar and antipolar phases are close in energy. In below, we demonstrate that 3L AlSb is an unusual tristable system that supports both ferroelectricity and antiferroelectricity.

\begin{figure}[ht]
    \centering
    \includegraphics[scale=0.45]{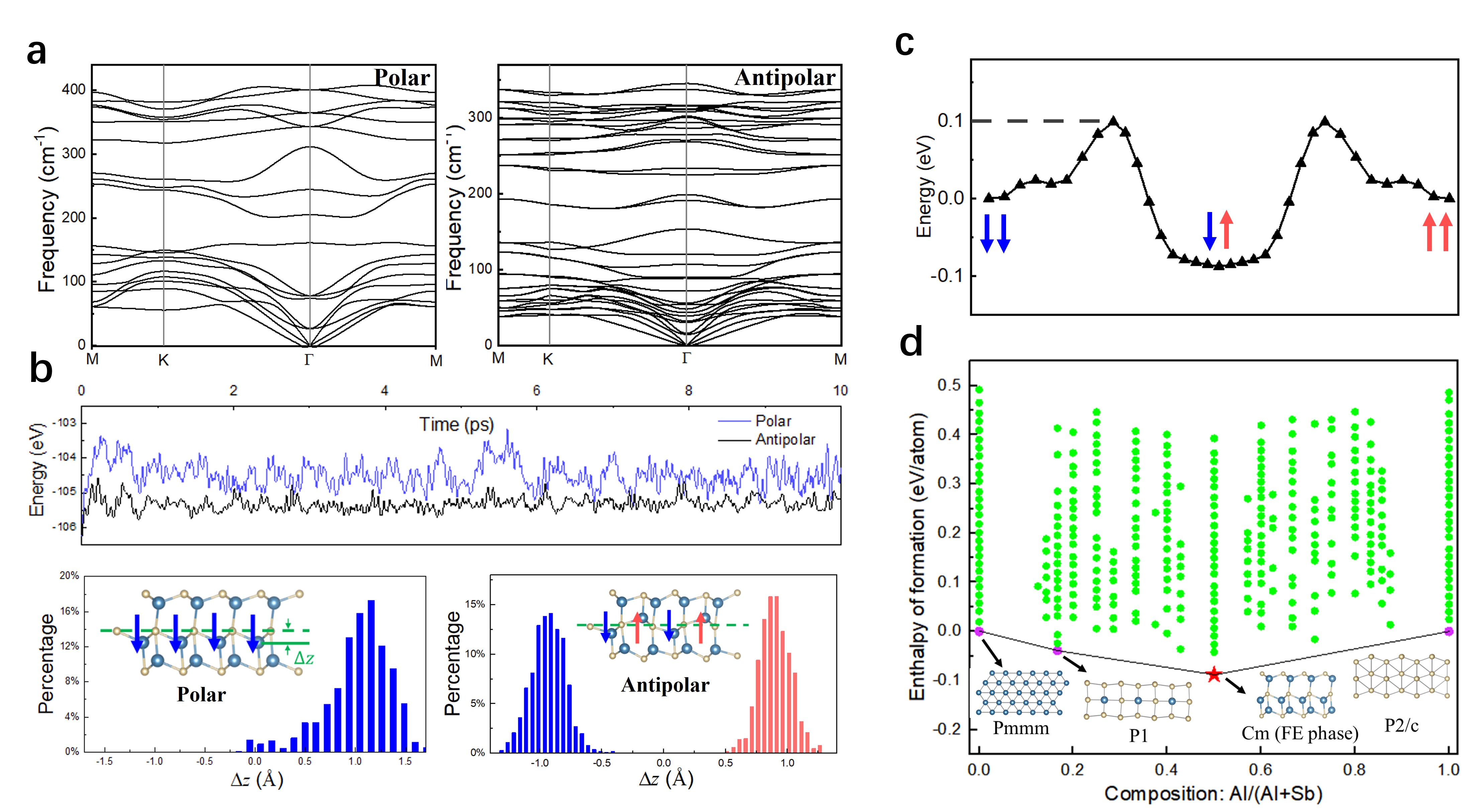}
    \caption{(a) Phonon dispersion relationships of 3L AlSb in the polar phase (left) and antipolar phase (right). (b) AIMD simulations. The top pannel shows the energy evolution as a function of time at 600~K. The bottom pannel shows the distribution of out-of-plane local displacements ($\Delta z$) of Al atoms in the central layer. (c) Minimum energy path obtained with NEB connecting the polar and antipolar phases. (d) Convex hull of Al$_x$Sb$_{1-x}$ from variable-composition evolutionary structure search.
    }
    \label{structure}
\end{figure}

Figure~\ref{structure}a presents the phonon spectra of 3L AlSb in polar and antipolar phase, respectively. Since the phonon spectra have no imaginary vibrational frequencies over the whole Brillouin zone, polar and nonpolar phases are dynamically stable and each locates at a local minimum of the potential energy surface. We perform AIMD simulations at elevated temperatures to check the structural stability against larger atomic distortions due to thermal fluctuations. The evolution of the total energy at 600~K during AIMD simulations is shown for both phases in Fig.~\ref{structure}b, revealing no sign of structural destruction or reconstruction. This serves as a strong evidence to corroborate the room-temperature stability of 3L AlSb. A defining feature of (anti)ferroelectricity is the polarization reversibility.
As depicted in Fig.~\ref{structure}c, the barrier separating the polar and antipolar phase obtained with the nudged elastic band (NEB) method is 0.1 eV that is lower than the barrier in conventional perovskite ferroelectrics such as PbTiO$_3$ (0.17 eV)~\cite{Huang22p144106}, indicating a switchable polarization by an external electric field. 
Therefore, 3L AlSb is a rare 2D material characterized by tristable and electrically switchable polarization states and thus hosts both ferroelectricity and antiferroelectricity. 

In addition, we perform a structural search using the variable-composition evolutionary algorithm as implemented in USPEX~\cite{Oganov06p244704, Lyakhov13p1172,Oganov11p227} with a 6-atom slab model confined within 9~\AA. Figure.~\ref{structure}d compiles the DFT formation enthalpies of all identified 2D crystals for Al$_{1-x}$Sb$_x$. We find that ferroelectric 3L AlSb has a convex hall distance of zero, further supporting its thermodynamic stability and synthesizability.

The emergence of $P_{\rm OP}$ in ferroelectric 3L AlSb can be understood by determining the electric free energy ($F$) of wurtzite AlSb under an open-circuit boundary condition (OCBC) that has $D=0$ where $D$ is the electric displacement. For an intermediate configuration $\lambda$ obtained by linear interpolating the ground-state polar configuration ($\lambda=1$) and the high-symmetry nonpolar configuration ($\lambda=0$, space group $P6_3/mmc$), the free energy $F(\lambda)$ under $D=0$ can be estimated as~\cite{Adhikari19p104101, Qiu21p064112}
\begin{equation}
F(\lambda) = U(\lambda) + \Omega(\lambda)\frac{1+\frac{1}{2}\chi_{\infty}(\lambda)}{\epsilon_0[1+\chi_{\infty} (\lambda)]^2}P^2(\lambda)
\label{free}
\end{equation}
where $U(\lambda)$, $P(\lambda)$, $\chi_{\infty}(\lambda)$, and $\Omega(\lambda)$ are the DFT total (internal) energy per unit cell, electric polarization, high-frequency dielectric permittivity along the polar direction, and the unit cell volume of AlSb at configuration $\lambda$, respectively, and $\epsilon_0$ is the vacuum permittivity. The internal energy $U(\lambda)$ becomes the electric free energy under short-circuit boundary condition (SCBC, $\mathcal{E}=0$) and the second term is the depolarization energy $f_d$ associated with the depolarization field under OCBC. All quantities required to evaluate $F(\lambda)$ are bulk values easily accessible via conventional DFT calculations. This analytical formation of $F(\lambda)$ has been used to understand the origin of hyperferroelectricity~\cite{Adhikari19p104101} in thin films under OCBC. 

\begin{figure}[ht]
    \centering
    \includegraphics[scale=0.5]{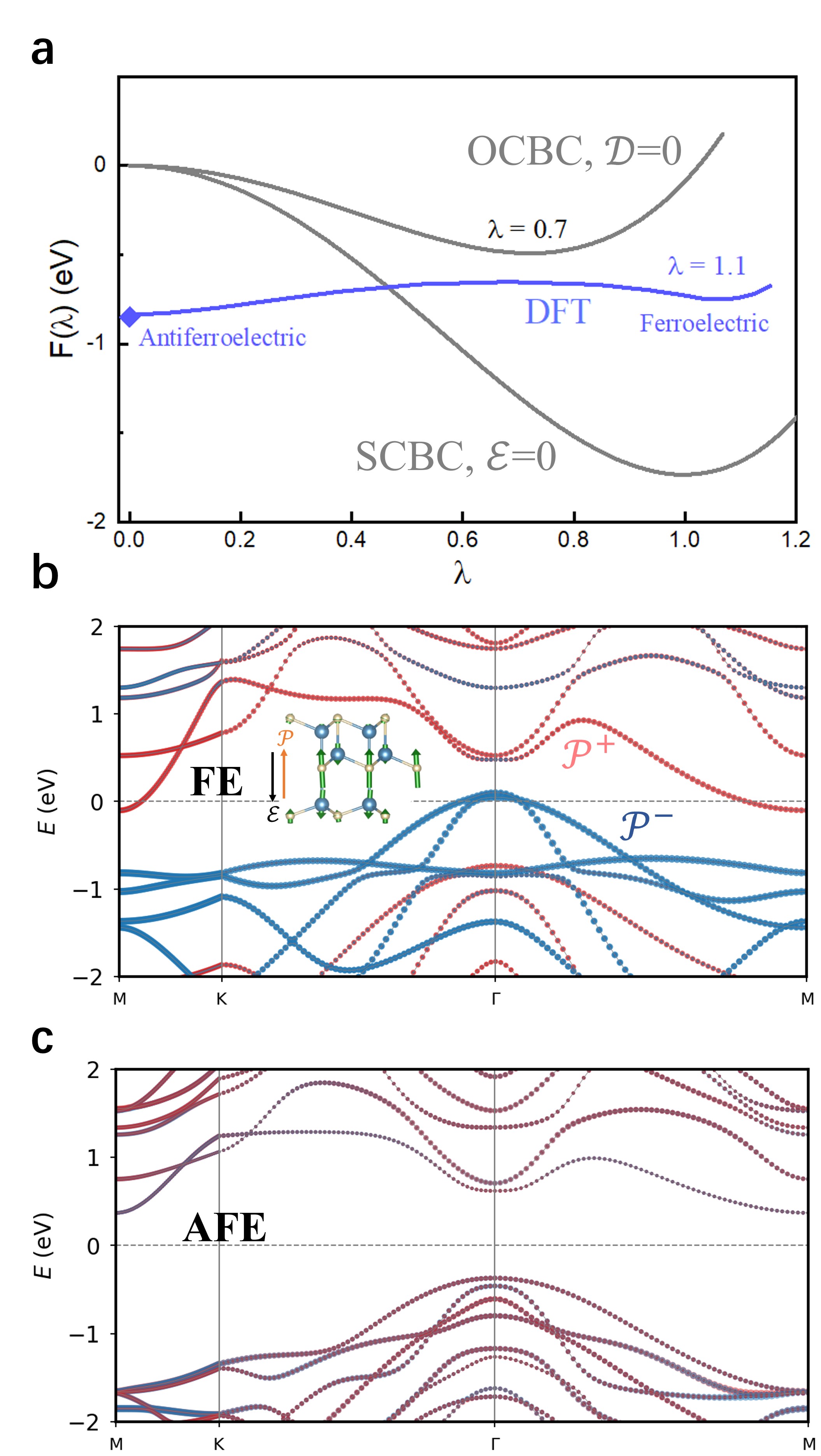}
    \caption{(a) Electric free energy $F(\lambda)$ of AlSb under SCBC ($E=0$) and OCBC ($D=0$). The DFT minimum energy pathway (blue line) connecting the ferroelectric (FE) and antiferroelectric (AFE) phases in 3L AlSb is plotted for comparison. Electronic band structures of (b) ferroelectric and (c) antiferroelectric 3L AlSb computed with eACBN0. The ferroelectric phase is a semimetal and the projected band structure has atomic orbital contributions from $P^-$ and $P^+$ surfaces colored in blue and red, respectively. The inset in (b) shows the atomic forces induced by an electric field applied against $P_{\rm OP}$, showing that nearly all atoms are affected by the applied field despite that the ferroelectric phase is a semimetal.
    }
    \label{ele}
\end{figure}

As shown in Fig.~\ref{ele}a, the potential well of $U(\lambda)$ is rather deep under SCBC. After introducing the depolarization effect under OCBC, the well becomes shallower and the ground state remains polar as  $F(\lambda)$ reaches the minimum at $\lambda=$0.7. It is noted that Eq.~\ref{free} does not consider the impact of surface reconstruction or the change in $\chi_{\infty}(\lambda)$ with reduced dimension. Nevertheless, the simple analytical model of Eq.~\ref{free} predicts that AlSb has a low-energy polar state under OCBC, resembling a hyperferroelectric~\cite{Garrity14p127601}.
We also plot the DFT energy profile for the ferroelectric-antiferroelectric transition in 3L AlSb in Fig.~\ref{ele}a. By comparing the analytical and DFT results, we suggest the surface reconstruction of 3L AlSb that has group-V anions becoming the outmost surface layers strongly stabilize the polar phase ($\lambda=1.1$), while the emergence of a low-energy antiferroelectric phase (not captured by the analytical model) is critical for the polarization switchability.

We now consider the electronic properties of 2L and 3L AlSb. Semilocal density functionals such as PBE often underestimate the band gap due to the remnant self-interaction error.  To obtain accurate electronic structures of 3L AlSb, we employ 
a newly developed pseudohybrid Hubbard density functional, extend Agapito–Cuetarolo–Buongiorno Nardelli (eACBN0)~\cite{Agapito13p165127, Lee20p043410, Tancogne-Dejean20p155117}. The eACBN0 function is a DFT+$U$+$V$ method with self-consistently computed  Hubbard $U$ ($V$) parameters that account for the onsite (intersite) Coulomb interactions, thus capable of capturing the local variations of Coulomb screening. Particularly for low-dimensional materials, eACBN0 yields better descriptions of the electronic structures than hybrid density functionals such as HSE06~\cite{Krukau06p224106} that assumes
fixed dielectric screening~\cite{Huang20p165157,Lee20p043410}.
Figure~\ref{ele}b-c presents the eACBN0 band structures for 3L AlSb in ferroelectric and antiferroelectric phases (see the comparison of eACBN0 and HSE06 band structures in Supporting Information). We find that the ferroelectric phase is a semimetal while the antiferroelectric phase is a semiconductor with a band gap of 0.7~eV. The semimetal nature of the ferroelectric phase is due to the built-in depolarization field that induces a band bending~\cite{Ke21p3387,Huang22p1440} such that the valence band
maximum (VBM) and the conduction band minimum  (CBM) are
dominated by the states of $P^-$ and $P^+$ surfaces, respectively (Fig.~\ref{ele}b). Moreover, we compute the
field-induced forces (Fig.~\ref{ele}b inset) and find that
nearly all atoms are affected by an applied field. This indciates the (semi)metallic ferroelectric 3AlSb remains electrically switchable, similar to 2D metallic WTe$_2$~\cite{Fei18p336}.
In contrast, the antiferroelectric phase has null depolarization field and the band gap is mostly determined by the hybridization of Al-$3p$ and Sb-$5p$ states. 
The strong coupling between the polarization state and the band gap in 3L AlSb enables intrinsically voltage-switchable metal-semiconductor transition~\cite{Duan21p2316}, a useful feature to realize on-off states for device applications.  

\begin{figure}[ht]
    \centering
    \includegraphics[scale=0.5]{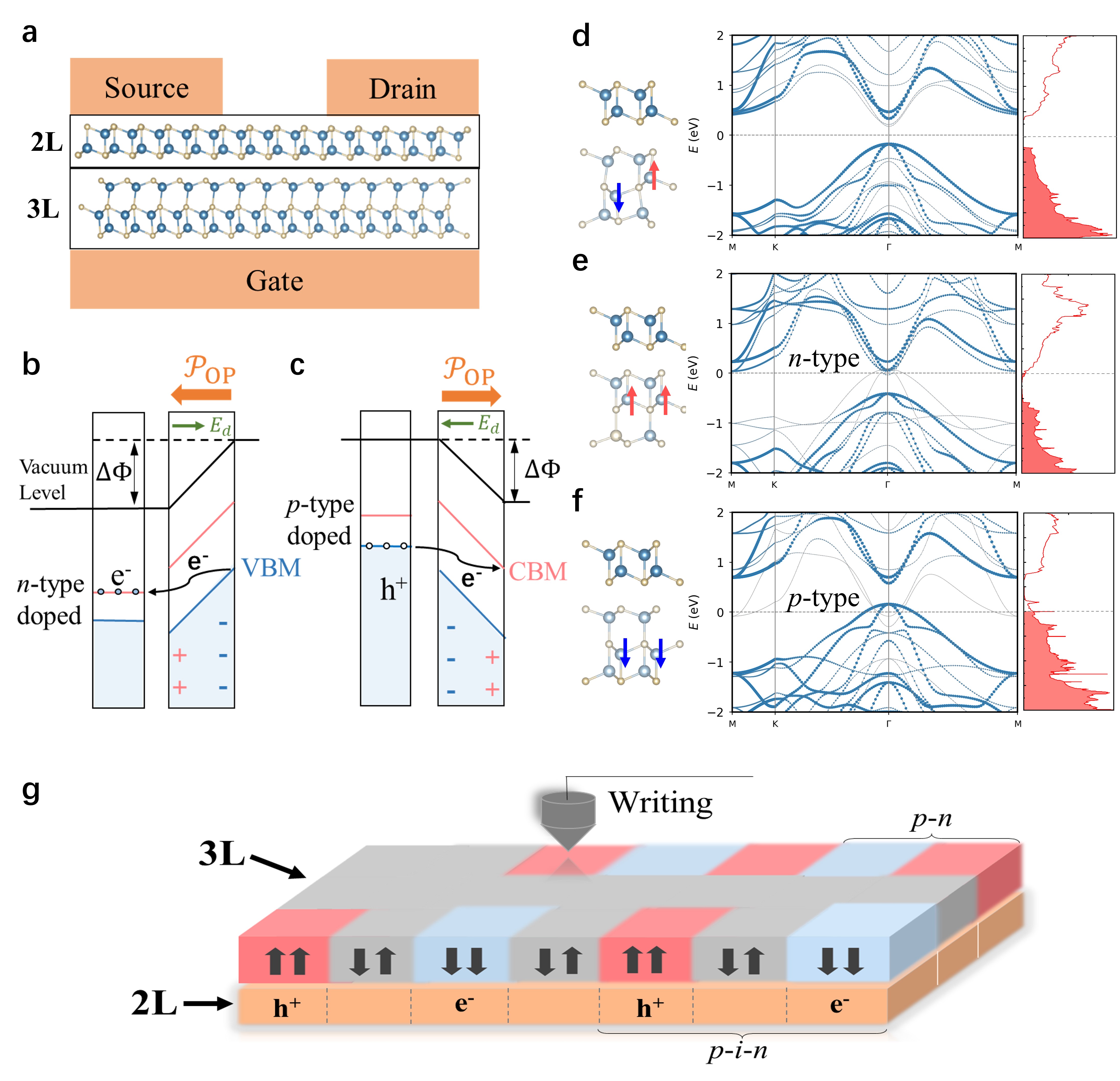}
    \caption{(a) Sechmatic of a 2D homojunction field effect transistor consisting of semiconducting 2L AlSb and tristable 3L AlSb. (b) Band bending diagrams of 2L-3L homojunction. The depolarization field $E_d$ in ferroelectric 3L AlSb creates a potential step $\Delta \Phi$ across the sheet. When $P_{\rm OP}$ points toward 2L AlSb, the high-energy electrons transfer from 3L to 2L, making 2L $n$-type doped. The polarization reversal will lead to a $p$-type doped 2L AlSb. Projected electronic band structures and density of states of 2L-3L homojunction with 3L adopting (d) antiferroelectric, (e) upward polarization, and (f) downward polarization, showing atomic orbital contributions from 2L AlSb. (g) Schematic of voltage-configurable multidomain-determined high-density $p$-$n$ and $p$-$i$-$n$ junction arrays.
    }
    \label{device}
\end{figure}

The ferroelectric field effect transistor (FeFET) comprising a semiconductor as the channel material and a ferroelectric as the gate insulator is an attractive architecture to realize low-power, high-speed, and high-density nonvolatile memory. Our DFT calculations show that 2L AlSb is a nonpolar semiconductor with a band gap of 0.5~eV. Taking advantage of the semiconducting property of 2L AlSb and the tristable polarization states affored by 3L AlSb, we propose a 2D homojunction FET using 3L AlSb as the gate insulator and 2L AlSb as the channel material (Fig.~\ref{device}a). The design based on a homojunction could simply the fabrication process and improve the device performance over the heterojucntion-based device by reducing interfacial defects.

We compute the eACBN0 band structures of a 2L-3L homojunction with 3L AlSb adopting different polarization states. The contributions from states of 2L AlSb are highlighted in the band structures to reveal the electrical properties of the channel. As shown in Fig.~\ref{device}d-f, the conductivity of 2L AlSb is readily regulated by the polarization state of 3L AlSb. Specifically, when the 3L AlSb adopts the antiferroelectric state, the channel consisting of 2L AlSb is a semicondcutor with a band gap of $\approx$0.6~eV (Fig.~\ref{device}d). When the polarization of 3L AlSb is switched toward 2L AlSb, the band structure of the homojunction reveals a $n$-type doped 2L AlSb (Fig.~\ref{device}e). This can be understood from the band diagram (right before the charge transfer) illustrated in Fig.~\ref{device}b. Because the VBM of 3L AlSb is higher in energy than the CBM of 2L AlSb, high-energy electrons in 3L AlSb naturally relax to the conduction bands of 2L AlSb, effectively $n$-type doping the channel. Finally, in the case where 3L AlSb has the polarization pointing away from 2L AlSb, the channel becomes hole doped as the electrons in 2L AlSb relax to the CBM of 3L AlSb that is lower in energy (Fig.~\ref{device}c). Therefore, the tristable polarization states of 3L AlSb create three resistance states of the channel, suitable for nonvolatile multistate functional applications. 

In addition, the nanoscale deterministic ferroelectric domain engineering can be employed to configure the 2L-3L homojunction into high-density $p$-$n$ junction arrays as well as $p$-$i$-$n$ junction arrays where the tristable polarization states of 3L AlSb control the carrier type and density in 2L AlSb, as shown in Fig.~\ref{device}g. The voltage-configurable multidomain pattern offers a platform to design energy-efficient, high-density synaptic electronics and neuromorphic systems.

In summary, we propose a strategy to obtain switchable 2D polar materials with promising compatibility with the main stream semiconductor industry. The depolarization field that is  
often considered detrimental to ferroelectric properties is used as a physical stressor to convert unswitchable bulk III-V semiconductors to 2D materials with reversible polarization. 
The delicate competition between the local polarization energy, the global depolarization energy, and the neighboring dipolar coupling in 2D gives rise to a thickness-sensitive phase competition.
The triatomic layer of AlSb is demonstrated to exhibit tristable polarization states thus hosting both ferroelectricity and antiferroelectricity. We have explored the functionalities of AlSb-based 2D homojunctions consisting of diatomic and triatomic layers and predicted the emergence of three distinct and nonvolatile resistance states characterized by different carrier type and density. The readily regulable doping by the tristable polarization states potentially enables facile fabrications of high-density periodic $p$-$n$ and $p$-$i$-$n$ junctions at the nanoscale for nanoelectric and optoelectronic devices.


\begin{acknowledgments}
C.K., S.L. acknowledge the supports from Westlake Education Foundation. The computational resource is provided by Westlake HPC Center.
\end{acknowledgments}



\newpage
\bibliography{SL}

\begin{thebibliography}{50}%
\makeatletter
\providecommand \@ifxundefined [1]{%
 \@ifx{#1\undefined}
}%
\providecommand \@ifnum [1]{%
 \ifnum #1\expandafter \@firstoftwo
 \else \expandafter \@secondoftwo
 \fi
}%
\providecommand \@ifx [1]{%
 \ifx #1\expandafter \@firstoftwo
 \else \expandafter \@secondoftwo
 \fi
}%
\providecommand \natexlab [1]{#1}%
\providecommand \enquote  [1]{``#1''}%
\providecommand \bibnamefont  [1]{#1}%
\providecommand \bibfnamefont [1]{#1}%
\providecommand \citenamefont [1]{#1}%
\providecommand \href@noop [0]{\@secondoftwo}%
\providecommand \href [0]{\begingroup \@sanitize@url \@href}%
\providecommand \@href[1]{\@@startlink{#1}\@@href}%
\providecommand \@@href[1]{\endgroup#1\@@endlink}%
\providecommand \@sanitize@url [0]{\catcode `\\12\catcode `\$12\catcode
  `\&12\catcode `\#12\catcode `\^12\catcode `\_12\catcode `\%12\relax}%
\providecommand \@@startlink[1]{}%
\providecommand \@@endlink[0]{}%
\providecommand \url  [0]{\begingroup\@sanitize@url \@url }%
\providecommand \@url [1]{\endgroup\@href {#1}{\urlprefix }}%
\providecommand \urlprefix  [0]{URL }%
\providecommand \Eprint [0]{\href }%
\providecommand \doibase [0]{https://doi.org/}%
\providecommand \selectlanguage [0]{\@gobble}%
\providecommand \bibinfo  [0]{\@secondoftwo}%
\providecommand \bibfield  [0]{\@secondoftwo}%
\providecommand \translation [1]{[#1]}%
\providecommand \BibitemOpen [0]{}%
\providecommand \bibitemStop [0]{}%
\providecommand \bibitemNoStop [0]{.\EOS\space}%
\providecommand \EOS [0]{\spacefactor3000\relax}%
\providecommand \BibitemShut  [1]{\csname bibitem#1\endcsname}%
\let\auto@bib@innerbib\@empty
\bibitem [{\citenamefont {Scott}(2007)}]{Scott07p954}%
  \BibitemOpen
  \bibfield  {author} {\bibinfo {author} {\bibfnamefont {J.~F.}\ \bibnamefont
  {Scott}},\ }\bibfield  {title} {\bibinfo {title} {Applications of modern
  ferroelectrics},\ }\href@noop {} {\bibfield  {journal} {\bibinfo  {journal}
  {Science}\ }\textbf {\bibinfo {volume} {315}},\ \bibinfo {pages} {954}
  (\bibinfo {year} {2007})}\BibitemShut {NoStop}%
\bibitem [{\citenamefont {Martin}\ and\ \citenamefont
  {Rappe}(2016)}]{Martin16p16087}%
  \BibitemOpen
  \bibfield  {author} {\bibinfo {author} {\bibfnamefont {L.~W.}\ \bibnamefont
  {Martin}}\ and\ \bibinfo {author} {\bibfnamefont {A.~M.}\ \bibnamefont
  {Rappe}},\ }\bibfield  {title} {\bibinfo {title} {Thin-film ferroelectric
  materials and their applications},\ }\href
  {https://doi.org/10.1038/natrevmats.2016.87} {\bibfield  {journal} {\bibinfo
  {journal} {Nat. Rev. Mater.}\ }\textbf {\bibinfo {volume} {2}},\ \bibinfo
  {pages} {16087} (\bibinfo {year} {2016})}\BibitemShut {NoStop}%
\bibitem [{\citenamefont {Mikolajick}\ \emph {et~al.}(2021)\citenamefont
  {Mikolajick}, \citenamefont {Slesazeck}, \citenamefont {Mulaosmanovic},
  \citenamefont {Park}, \citenamefont {Fichtner}, \citenamefont {Lomenzo},
  \citenamefont {Hoffmann},\ and\ \citenamefont
  {Schroeder}}]{Mikolajick21p100901}%
  \BibitemOpen
  \bibfield  {author} {\bibinfo {author} {\bibfnamefont {T.}~\bibnamefont
  {Mikolajick}}, \bibinfo {author} {\bibfnamefont {S.}~\bibnamefont
  {Slesazeck}}, \bibinfo {author} {\bibfnamefont {H.}~\bibnamefont
  {Mulaosmanovic}}, \bibinfo {author} {\bibfnamefont {M.~H.}\ \bibnamefont
  {Park}}, \bibinfo {author} {\bibfnamefont {S.}~\bibnamefont {Fichtner}},
  \bibinfo {author} {\bibfnamefont {P.~D.}\ \bibnamefont {Lomenzo}}, \bibinfo
  {author} {\bibfnamefont {M.}~\bibnamefont {Hoffmann}},\ and\ \bibinfo
  {author} {\bibfnamefont {U.}~\bibnamefont {Schroeder}},\ }\bibfield  {title}
  {\bibinfo {title} {Next generation ferroelectric materials for semiconductor
  process integration and their applications},\ }\href
  {https://doi.org/10.1063/5.0037617} {\bibfield  {journal} {\bibinfo
  {journal} {J. Appl. Phys.}\ }\textbf {\bibinfo {volume} {129}},\ \bibinfo
  {pages} {100901} (\bibinfo {year} {2021})}\BibitemShut {NoStop}%
\bibitem [{\citenamefont {Wu}\ and\ \citenamefont {Jena}(2018)}]{Wu18pe1365}%
  \BibitemOpen
  \bibfield  {author} {\bibinfo {author} {\bibfnamefont {M.}~\bibnamefont
  {Wu}}\ and\ \bibinfo {author} {\bibfnamefont {P.}~\bibnamefont {Jena}},\
  }\bibfield  {title} {\bibinfo {title} {The rise of two-dimensional van der
  waals ferroelectrics},\ }\href {https://doi.org/10.1002/wcms.1365} {\bibfield
   {journal} {\bibinfo  {journal} {{WIREs} Comput. Mol. Sci.}\ }\textbf
  {\bibinfo {volume} {8}},\ \bibinfo {pages} {e1365} (\bibinfo {year}
  {2018})}\BibitemShut {NoStop}%
\bibitem [{\citenamefont {Guan}\ \emph {et~al.}(2019)\citenamefont {Guan},
  \citenamefont {Hu}, \citenamefont {Shen}, \citenamefont {Xiang},
  \citenamefont {Zhong}, \citenamefont {Chu},\ and\ \citenamefont
  {Duan}}]{Guan19p1900818}%
  \BibitemOpen
  \bibfield  {author} {\bibinfo {author} {\bibfnamefont {Z.}~\bibnamefont
  {Guan}}, \bibinfo {author} {\bibfnamefont {H.}~\bibnamefont {Hu}}, \bibinfo
  {author} {\bibfnamefont {X.}~\bibnamefont {Shen}}, \bibinfo {author}
  {\bibfnamefont {P.}~\bibnamefont {Xiang}}, \bibinfo {author} {\bibfnamefont
  {N.}~\bibnamefont {Zhong}}, \bibinfo {author} {\bibfnamefont
  {J.}~\bibnamefont {Chu}},\ and\ \bibinfo {author} {\bibfnamefont
  {C.}~\bibnamefont {Duan}},\ }\bibfield  {title} {\bibinfo {title} {Recent
  progress in two-dimensional ferroelectric materials},\ }\href
  {https://doi.org/10.1002/aelm.201900818} {\bibfield  {journal} {\bibinfo
  {journal} {Adv. Electron. Mater.}\ }\textbf {\bibinfo {volume} {6}},\
  \bibinfo {pages} {1900818} (\bibinfo {year} {2019})}\BibitemShut {NoStop}%
\bibitem [{\citenamefont {Kruse}\ \emph {et~al.}(2022)\citenamefont {Kruse},
  \citenamefont {Petralanda}, \citenamefont {Gjerding}, \citenamefont
  {Jacobsen}, \citenamefont {Thygesen},\ and\ \citenamefont
  {Olsen}}]{Kruse22arxiv}%
  \BibitemOpen
  \bibfield  {author} {\bibinfo {author} {\bibfnamefont {M.}~\bibnamefont
  {Kruse}}, \bibinfo {author} {\bibfnamefont {U.}~\bibnamefont {Petralanda}},
  \bibinfo {author} {\bibfnamefont {M.~N.}\ \bibnamefont {Gjerding}}, \bibinfo
  {author} {\bibfnamefont {K.~W.}\ \bibnamefont {Jacobsen}}, \bibinfo {author}
  {\bibfnamefont {K.~S.}\ \bibnamefont {Thygesen}},\ and\ \bibinfo {author}
  {\bibfnamefont {T.}~\bibnamefont {Olsen}},\ }\bibfield  {title} {\bibinfo
  {title} {Two-dimensional ferroelectrics from high throughput computational
  screening},\ }\href {https://doi.org/10.48550/ARXIV.2209.13911} {\bibfield
  {journal} {\bibinfo  {journal} {arXiv}\ ,\ \bibinfo {pages} {2102.11508}}
  (\bibinfo {year} {2022})}\BibitemShut {NoStop}%
\bibitem [{\citenamefont {Liu}\ \emph {et~al.}(2016)\citenamefont {Liu},
  \citenamefont {You}, \citenamefont {Seyler}, \citenamefont {Li},
  \citenamefont {Yu}, \citenamefont {Lin}, \citenamefont {Wang}, \citenamefont
  {Zhou}, \citenamefont {Wang}, \citenamefont {He}, \citenamefont {Pantelides},
  \citenamefont {Zhou}, \citenamefont {Sharma}, \citenamefont {Xu},
  \citenamefont {Ajayan}, \citenamefont {Wang},\ and\ \citenamefont
  {Liu}}]{Liu16p12357}%
  \BibitemOpen
  \bibfield  {author} {\bibinfo {author} {\bibfnamefont {F.}~\bibnamefont
  {Liu}}, \bibinfo {author} {\bibfnamefont {L.}~\bibnamefont {You}}, \bibinfo
  {author} {\bibfnamefont {K.~L.}\ \bibnamefont {Seyler}}, \bibinfo {author}
  {\bibfnamefont {X.}~\bibnamefont {Li}}, \bibinfo {author} {\bibfnamefont
  {P.}~\bibnamefont {Yu}}, \bibinfo {author} {\bibfnamefont {J.}~\bibnamefont
  {Lin}}, \bibinfo {author} {\bibfnamefont {X.}~\bibnamefont {Wang}}, \bibinfo
  {author} {\bibfnamefont {J.}~\bibnamefont {Zhou}}, \bibinfo {author}
  {\bibfnamefont {H.}~\bibnamefont {Wang}}, \bibinfo {author} {\bibfnamefont
  {H.}~\bibnamefont {He}}, \bibinfo {author} {\bibfnamefont {S.~T.}\
  \bibnamefont {Pantelides}}, \bibinfo {author} {\bibfnamefont
  {W.}~\bibnamefont {Zhou}}, \bibinfo {author} {\bibfnamefont {P.}~\bibnamefont
  {Sharma}}, \bibinfo {author} {\bibfnamefont {X.}~\bibnamefont {Xu}}, \bibinfo
  {author} {\bibfnamefont {P.~M.}\ \bibnamefont {Ajayan}}, \bibinfo {author}
  {\bibfnamefont {J.}~\bibnamefont {Wang}},\ and\ \bibinfo {author}
  {\bibfnamefont {Z.}~\bibnamefont {Liu}},\ }\bibfield  {title} {\bibinfo
  {title} {Room-temperature ferroelectricity in {CuInP$_2$S$_6$} ultrathin
  flakes},\ }\href {https://doi.org/10.1038/ncomms12357} {\bibfield  {journal}
  {\bibinfo  {journal} {Nat. Commun.}\ }\textbf {\bibinfo {volume} {7}},\
  \bibinfo {pages} {12357} (\bibinfo {year} {2016})}\BibitemShut {NoStop}%
\bibitem [{\citenamefont {Ding}\ \emph {et~al.}(2017)\citenamefont {Ding},
  \citenamefont {Zhu}, \citenamefont {Wang}, \citenamefont {Gao}, \citenamefont
  {Xiao}, \citenamefont {Gu}, \citenamefont {Zhang},\ and\ \citenamefont
  {Zhu}}]{Ding17p14956}%
  \BibitemOpen
  \bibfield  {author} {\bibinfo {author} {\bibfnamefont {W.}~\bibnamefont
  {Ding}}, \bibinfo {author} {\bibfnamefont {J.}~\bibnamefont {Zhu}}, \bibinfo
  {author} {\bibfnamefont {Z.}~\bibnamefont {Wang}}, \bibinfo {author}
  {\bibfnamefont {Y.}~\bibnamefont {Gao}}, \bibinfo {author} {\bibfnamefont
  {D.}~\bibnamefont {Xiao}}, \bibinfo {author} {\bibfnamefont {Y.}~\bibnamefont
  {Gu}}, \bibinfo {author} {\bibfnamefont {Z.}~\bibnamefont {Zhang}},\ and\
  \bibinfo {author} {\bibfnamefont {W.}~\bibnamefont {Zhu}},\ }\bibfield
  {title} {\bibinfo {title} {Prediction of intrinsic two-dimensional
  ferroelectrics in {In$_2$Se$_3$} and other {III$_2$-VI$_3$} van der waals
  materials},\ }\href@noop {} {\bibfield  {journal} {\bibinfo  {journal} {Nat.
  Commun.}\ }\textbf {\bibinfo {volume} {8}},\ \bibinfo {pages} {14956}
  (\bibinfo {year} {2017})}\BibitemShut {NoStop}%
\bibitem [{\citenamefont {Zhou}\ \emph {et~al.}(2017)\citenamefont {Zhou},
  \citenamefont {Wu}, \citenamefont {Zhu}, \citenamefont {Cho}, \citenamefont
  {He}, \citenamefont {Yang}, \citenamefont {Herrera}, \citenamefont {Chu},
  \citenamefont {Han}, \citenamefont {Downer}, \citenamefont {Peng},\ and\
  \citenamefont {Lai}}]{Zhou17p5508}%
  \BibitemOpen
  \bibfield  {author} {\bibinfo {author} {\bibfnamefont {Y.}~\bibnamefont
  {Zhou}}, \bibinfo {author} {\bibfnamefont {D.}~\bibnamefont {Wu}}, \bibinfo
  {author} {\bibfnamefont {Y.}~\bibnamefont {Zhu}}, \bibinfo {author}
  {\bibfnamefont {Y.}~\bibnamefont {Cho}}, \bibinfo {author} {\bibfnamefont
  {Q.}~\bibnamefont {He}}, \bibinfo {author} {\bibfnamefont {X.}~\bibnamefont
  {Yang}}, \bibinfo {author} {\bibfnamefont {K.}~\bibnamefont {Herrera}},
  \bibinfo {author} {\bibfnamefont {Z.}~\bibnamefont {Chu}}, \bibinfo {author}
  {\bibfnamefont {Y.}~\bibnamefont {Han}}, \bibinfo {author} {\bibfnamefont
  {M.~C.}\ \bibnamefont {Downer}}, \bibinfo {author} {\bibfnamefont
  {H.}~\bibnamefont {Peng}},\ and\ \bibinfo {author} {\bibfnamefont
  {K.}~\bibnamefont {Lai}},\ }\bibfield  {title} {\bibinfo {title}
  {Out-of-plane piezoelectricity and ferroelectricity in layered
  {$\alpha$-In$_2$Se$_3$} nanoflakes},\ }\href
  {https://doi.org/10.1021/acs.nanolett.7b02198} {\bibfield  {journal}
  {\bibinfo  {journal} {Nano Lett.}\ }\textbf {\bibinfo {volume} {17}},\
  \bibinfo {pages} {5508} (\bibinfo {year} {2017})}\BibitemShut {NoStop}%
\bibitem [{\citenamefont {Cui}\ \emph {et~al.}(2018)\citenamefont {Cui},
  \citenamefont {Hu}, \citenamefont {Yan}, \citenamefont {Addiego},
  \citenamefont {Gao}, \citenamefont {Wang}, \citenamefont {Wang},
  \citenamefont {Li}, \citenamefont {Cheng}, \citenamefont {Li}, \citenamefont
  {Zhang}, \citenamefont {Alshareef}, \citenamefont {Wu}, \citenamefont {Zhu},
  \citenamefont {Pan},\ and\ \citenamefont {Li}}]{Cui18p1253}%
  \BibitemOpen
  \bibfield  {author} {\bibinfo {author} {\bibfnamefont {C.}~\bibnamefont
  {Cui}}, \bibinfo {author} {\bibfnamefont {W.-J.}\ \bibnamefont {Hu}},
  \bibinfo {author} {\bibfnamefont {X.}~\bibnamefont {Yan}}, \bibinfo {author}
  {\bibfnamefont {C.}~\bibnamefont {Addiego}}, \bibinfo {author} {\bibfnamefont
  {W.}~\bibnamefont {Gao}}, \bibinfo {author} {\bibfnamefont {Y.}~\bibnamefont
  {Wang}}, \bibinfo {author} {\bibfnamefont {Z.}~\bibnamefont {Wang}}, \bibinfo
  {author} {\bibfnamefont {L.}~\bibnamefont {Li}}, \bibinfo {author}
  {\bibfnamefont {Y.}~\bibnamefont {Cheng}}, \bibinfo {author} {\bibfnamefont
  {P.}~\bibnamefont {Li}}, \bibinfo {author} {\bibfnamefont {X.}~\bibnamefont
  {Zhang}}, \bibinfo {author} {\bibfnamefont {H.~N.}\ \bibnamefont
  {Alshareef}}, \bibinfo {author} {\bibfnamefont {T.}~\bibnamefont {Wu}},
  \bibinfo {author} {\bibfnamefont {W.}~\bibnamefont {Zhu}}, \bibinfo {author}
  {\bibfnamefont {X.}~\bibnamefont {Pan}},\ and\ \bibinfo {author}
  {\bibfnamefont {L.-J.}\ \bibnamefont {Li}},\ }\bibfield  {title} {\bibinfo
  {title} {Intercorrelated in-plane and out-of-plane ferroelectricity in
  ultrathin two-dimensional layered semiconductor {In$_2$Se$_3$}},\ }\href
  {https://doi.org/10.1021/acs.nanolett.7b04852} {\bibfield  {journal}
  {\bibinfo  {journal} {Nano Lett.}\ }\textbf {\bibinfo {volume} {18}},\
  \bibinfo {pages} {1253} (\bibinfo {year} {2018})}\BibitemShut {NoStop}%
\bibitem [{\citenamefont {Xiao}\ \emph {et~al.}(2018)\citenamefont {Xiao},
  \citenamefont {Zhu}, \citenamefont {Wang}, \citenamefont {Feng},
  \citenamefont {Hu}, \citenamefont {Dasgupta}, \citenamefont {Han},
  \citenamefont {Wang}, \citenamefont {Muller}, \citenamefont {Martin},
  \citenamefont {Hu},\ and\ \citenamefont {Zhang}}]{Xiao18p227601}%
  \BibitemOpen
  \bibfield  {author} {\bibinfo {author} {\bibfnamefont {J.}~\bibnamefont
  {Xiao}}, \bibinfo {author} {\bibfnamefont {H.}~\bibnamefont {Zhu}}, \bibinfo
  {author} {\bibfnamefont {Y.}~\bibnamefont {Wang}}, \bibinfo {author}
  {\bibfnamefont {W.}~\bibnamefont {Feng}}, \bibinfo {author} {\bibfnamefont
  {Y.}~\bibnamefont {Hu}}, \bibinfo {author} {\bibfnamefont {A.}~\bibnamefont
  {Dasgupta}}, \bibinfo {author} {\bibfnamefont {Y.}~\bibnamefont {Han}},
  \bibinfo {author} {\bibfnamefont {Y.}~\bibnamefont {Wang}}, \bibinfo {author}
  {\bibfnamefont {D.~A.}\ \bibnamefont {Muller}}, \bibinfo {author}
  {\bibfnamefont {L.~W.}\ \bibnamefont {Martin}}, \bibinfo {author}
  {\bibfnamefont {P.}~\bibnamefont {Hu}},\ and\ \bibinfo {author}
  {\bibfnamefont {X.}~\bibnamefont {Zhang}},\ }\bibfield  {title} {\bibinfo
  {title} {Intrinsic two-dimensional ferroelectricity with dipole locking},\
  }\href {https://doi.org/10.1103/PhysRevLett.120.227601} {\bibfield  {journal}
  {\bibinfo  {journal} {Phys. Rev. Lett.}\ }\textbf {\bibinfo {volume} {120}},\
  \bibinfo {pages} {227601} (\bibinfo {year} {2018})}\BibitemShut {NoStop}%
\bibitem [{\citenamefont {Poh}\ \emph {et~al.}(2018)\citenamefont {Poh},
  \citenamefont {Tan}, \citenamefont {Wang}, \citenamefont {Song},
  \citenamefont {Abidi}, \citenamefont {Zhao}, \citenamefont {Dan},
  \citenamefont {Chen}, \citenamefont {Luo}, \citenamefont {Pennycook},
  \citenamefont {Neto},\ and\ \citenamefont {Loh}}]{Poh18p6340}%
  \BibitemOpen
  \bibfield  {author} {\bibinfo {author} {\bibfnamefont {S.~M.}\ \bibnamefont
  {Poh}}, \bibinfo {author} {\bibfnamefont {S.~J.~R.}\ \bibnamefont {Tan}},
  \bibinfo {author} {\bibfnamefont {H.}~\bibnamefont {Wang}}, \bibinfo {author}
  {\bibfnamefont {P.}~\bibnamefont {Song}}, \bibinfo {author} {\bibfnamefont
  {I.~H.}\ \bibnamefont {Abidi}}, \bibinfo {author} {\bibfnamefont
  {X.}~\bibnamefont {Zhao}}, \bibinfo {author} {\bibfnamefont {J.}~\bibnamefont
  {Dan}}, \bibinfo {author} {\bibfnamefont {J.}~\bibnamefont {Chen}}, \bibinfo
  {author} {\bibfnamefont {Z.}~\bibnamefont {Luo}}, \bibinfo {author}
  {\bibfnamefont {S.~J.}\ \bibnamefont {Pennycook}}, \bibinfo {author}
  {\bibfnamefont {A.~H.~C.}\ \bibnamefont {Neto}},\ and\ \bibinfo {author}
  {\bibfnamefont {K.~P.}\ \bibnamefont {Loh}},\ }\bibfield  {title} {\bibinfo
  {title} {Molecular-beam epitaxy of two-dimensional {In$_2$Se$_3$} and its
  giant electroresistance switching in ferroresistive memory junction},\ }\href
  {https://doi.org/10.1021/acs.nanolett.8b02688} {\bibfield  {journal}
  {\bibinfo  {journal} {Nano Lett.}\ }\textbf {\bibinfo {volume} {18}},\
  \bibinfo {pages} {6340} (\bibinfo {year} {2018})}\BibitemShut {NoStop}%
\bibitem [{\citenamefont {Wan}\ \emph {et~al.}(2018)\citenamefont {Wan},
  \citenamefont {Li}, \citenamefont {Li}, \citenamefont {Mao}, \citenamefont
  {Zhu},\ and\ \citenamefont {Zeng}}]{Wan18p14885}%
  \BibitemOpen
  \bibfield  {author} {\bibinfo {author} {\bibfnamefont {S.}~\bibnamefont
  {Wan}}, \bibinfo {author} {\bibfnamefont {Y.}~\bibnamefont {Li}}, \bibinfo
  {author} {\bibfnamefont {W.}~\bibnamefont {Li}}, \bibinfo {author}
  {\bibfnamefont {X.}~\bibnamefont {Mao}}, \bibinfo {author} {\bibfnamefont
  {W.}~\bibnamefont {Zhu}},\ and\ \bibinfo {author} {\bibfnamefont
  {H.}~\bibnamefont {Zeng}},\ }\bibfield  {title} {\bibinfo {title}
  {Room-temperature ferroelectricity and a switchable diode effect in
  two-dimensional {$\alpha$-In$_2$Se$_3$} thin layers},\ }\href
  {https://doi.org/10.1039/c8nr04422h} {\bibfield  {journal} {\bibinfo
  {journal} {Nanoscale}\ }\textbf {\bibinfo {volume} {10}},\ \bibinfo {pages}
  {14885} (\bibinfo {year} {2018})}\BibitemShut {NoStop}%
\bibitem [{\citenamefont {Yuan}\ \emph {et~al.}(2019)\citenamefont {Yuan},
  \citenamefont {Luo}, \citenamefont {Chan}, \citenamefont {Xiao},
  \citenamefont {Dai}, \citenamefont {Xie},\ and\ \citenamefont
  {Hao}}]{Yuan19p1775}%
  \BibitemOpen
  \bibfield  {author} {\bibinfo {author} {\bibfnamefont {S.}~\bibnamefont
  {Yuan}}, \bibinfo {author} {\bibfnamefont {X.}~\bibnamefont {Luo}}, \bibinfo
  {author} {\bibfnamefont {H.~L.}\ \bibnamefont {Chan}}, \bibinfo {author}
  {\bibfnamefont {C.}~\bibnamefont {Xiao}}, \bibinfo {author} {\bibfnamefont
  {Y.}~\bibnamefont {Dai}}, \bibinfo {author} {\bibfnamefont {M.}~\bibnamefont
  {Xie}},\ and\ \bibinfo {author} {\bibfnamefont {J.}~\bibnamefont {Hao}},\
  }\bibfield  {title} {\bibinfo {title} {Room-temperature ferroelectricity in
  {MoTe$_2$} down to the atomic monolayer limit},\ }\href
  {https://doi.org/10.1038/s41467-019-09669-x} {\bibfield  {journal} {\bibinfo
  {journal} {Nat. Commun.}\ }\textbf {\bibinfo {volume} {10}},\ \bibinfo
  {pages} {1775} (\bibinfo {year} {2019})}\BibitemShut {NoStop}%
\bibitem [{\citenamefont {Fei}\ \emph {et~al.}(2018)\citenamefont {Fei},
  \citenamefont {Zhao}, \citenamefont {Palomaki}, \citenamefont {Sun},
  \citenamefont {Miller}, \citenamefont {Zhao}, \citenamefont {Yan},
  \citenamefont {Xu},\ and\ \citenamefont {Cobden}}]{Fei18p336}%
  \BibitemOpen
  \bibfield  {author} {\bibinfo {author} {\bibfnamefont {Z.}~\bibnamefont
  {Fei}}, \bibinfo {author} {\bibfnamefont {W.}~\bibnamefont {Zhao}}, \bibinfo
  {author} {\bibfnamefont {T.~A.}\ \bibnamefont {Palomaki}}, \bibinfo {author}
  {\bibfnamefont {B.}~\bibnamefont {Sun}}, \bibinfo {author} {\bibfnamefont
  {M.~K.}\ \bibnamefont {Miller}}, \bibinfo {author} {\bibfnamefont
  {Z.}~\bibnamefont {Zhao}}, \bibinfo {author} {\bibfnamefont {J.}~\bibnamefont
  {Yan}}, \bibinfo {author} {\bibfnamefont {X.}~\bibnamefont {Xu}},\ and\
  \bibinfo {author} {\bibfnamefont {D.~H.}\ \bibnamefont {Cobden}},\ }\bibfield
   {title} {\bibinfo {title} {Ferroelectric switching of a two-dimensional
  metal},\ }\href {https://doi.org/10.1038/s41586-018-0336-3} {\bibfield
  {journal} {\bibinfo  {journal} {Nature}\ }\textbf {\bibinfo {volume} {560}},\
  \bibinfo {pages} {336} (\bibinfo {year} {2018})}\BibitemShut {NoStop}%
\bibitem [{\citenamefont {Ferri}\ \emph {et~al.}(2021)\citenamefont {Ferri},
  \citenamefont {Bachu}, \citenamefont {Zhu}, \citenamefont {Imperatore},
  \citenamefont {Hayden}, \citenamefont {Alem}, \citenamefont {Giebink},
  \citenamefont {Trolier-McKinstry},\ and\ \citenamefont
  {Maria}}]{Ferri21p044101}%
  \BibitemOpen
  \bibfield  {author} {\bibinfo {author} {\bibfnamefont {K.}~\bibnamefont
  {Ferri}}, \bibinfo {author} {\bibfnamefont {S.}~\bibnamefont {Bachu}},
  \bibinfo {author} {\bibfnamefont {W.}~\bibnamefont {Zhu}}, \bibinfo {author}
  {\bibfnamefont {M.}~\bibnamefont {Imperatore}}, \bibinfo {author}
  {\bibfnamefont {J.}~\bibnamefont {Hayden}}, \bibinfo {author} {\bibfnamefont
  {N.}~\bibnamefont {Alem}}, \bibinfo {author} {\bibfnamefont {N.}~\bibnamefont
  {Giebink}}, \bibinfo {author} {\bibfnamefont {S.}~\bibnamefont
  {Trolier-McKinstry}},\ and\ \bibinfo {author} {\bibfnamefont {J.-P.}\
  \bibnamefont {Maria}},\ }\bibfield  {title} {\bibinfo {title} {Ferroelectrics
  everywhere: Ferroelectricity in magnesium substituted zinc oxide thin
  films},\ }\href {https://doi.org/10.1063/5.0053755} {\bibfield  {journal}
  {\bibinfo  {journal} {J. Appl. Phys.}\ }\textbf {\bibinfo {volume} {130}},\
  \bibinfo {pages} {044101} (\bibinfo {year} {2021})}\BibitemShut {NoStop}%
\bibitem [{\citenamefont {Fichtner}\ \emph {et~al.}(2019)\citenamefont
  {Fichtner}, \citenamefont {Wolff}, \citenamefont {Lofink}, \citenamefont
  {Kienle},\ and\ \citenamefont {Wagner}}]{Fichtner19p114103}%
  \BibitemOpen
  \bibfield  {author} {\bibinfo {author} {\bibfnamefont {S.}~\bibnamefont
  {Fichtner}}, \bibinfo {author} {\bibfnamefont {N.}~\bibnamefont {Wolff}},
  \bibinfo {author} {\bibfnamefont {F.}~\bibnamefont {Lofink}}, \bibinfo
  {author} {\bibfnamefont {L.}~\bibnamefont {Kienle}},\ and\ \bibinfo {author}
  {\bibfnamefont {B.}~\bibnamefont {Wagner}},\ }\bibfield  {title} {\bibinfo
  {title} {{AlScN}: A {III-V} semiconductor based ferroelectric},\ }\href
  {https://doi.org/10.1063/1.5084945} {\bibfield  {journal} {\bibinfo
  {journal} {J. Appl. Phys.}\ }\textbf {\bibinfo {volume} {125}},\ \bibinfo
  {pages} {114103} (\bibinfo {year} {2019})}\BibitemShut {NoStop}%
\bibitem [{\citenamefont {Jeong}\ \emph {et~al.}(2022)\citenamefont {Jeong},
  \citenamefont {Kim}, \citenamefont {Kim}, \citenamefont {Geum}, \citenamefont
  {Kim}, \citenamefont {Jo}, \citenamefont {Jeong}, \citenamefont {Park},
  \citenamefont {Jang}, \citenamefont {Choi}, \citenamefont {Kwon},\ and\
  \citenamefont {Kim}}]{jeong22p9031}%
  \BibitemOpen
  \bibfield  {author} {\bibinfo {author} {\bibfnamefont {J.}~\bibnamefont
  {Jeong}}, \bibinfo {author} {\bibfnamefont {S.~K.}\ \bibnamefont {Kim}},
  \bibinfo {author} {\bibfnamefont {J.}~\bibnamefont {Kim}}, \bibinfo {author}
  {\bibfnamefont {D.-M.}\ \bibnamefont {Geum}}, \bibinfo {author}
  {\bibfnamefont {D.}~\bibnamefont {Kim}}, \bibinfo {author} {\bibfnamefont
  {E.}~\bibnamefont {Jo}}, \bibinfo {author} {\bibfnamefont {H.}~\bibnamefont
  {Jeong}}, \bibinfo {author} {\bibfnamefont {J.}~\bibnamefont {Park}},
  \bibinfo {author} {\bibfnamefont {J.-H.}\ \bibnamefont {Jang}}, \bibinfo
  {author} {\bibfnamefont {S.}~\bibnamefont {Choi}}, \bibinfo {author}
  {\bibfnamefont {I.}~\bibnamefont {Kwon}},\ and\ \bibinfo {author}
  {\bibfnamefont {S.}~\bibnamefont {Kim}},\ }\bibfield  {title} {\bibinfo
  {title} {Heterogeneous and monolithic 3d integration of {III–V}-based radio
  frequency devices on {Si} {CMOS} circuits},\ }\href@noop {} {\bibfield
  {journal} {\bibinfo  {journal} {{ACS} Nano}\ }\textbf {\bibinfo {volume}
  {16}},\ \bibinfo {pages} {9031} (\bibinfo {year} {2022})}\BibitemShut
  {NoStop}%
\bibitem [{\citenamefont {Chen}\ \emph {et~al.}(2016)\citenamefont {Chen},
  \citenamefont {Li}, \citenamefont {Wu}, \citenamefont {Jiang}, \citenamefont
  {Tang}, \citenamefont {Shutts}, \citenamefont {Elliott}, \citenamefont
  {Sobiesierski}, \citenamefont {Seeds}, \citenamefont {Ross}, \citenamefont
  {Smowton},\ and\ \citenamefont {Liu}}]{chen16p307}%
  \BibitemOpen
  \bibfield  {author} {\bibinfo {author} {\bibfnamefont {S.}~\bibnamefont
  {Chen}}, \bibinfo {author} {\bibfnamefont {W.}~\bibnamefont {Li}}, \bibinfo
  {author} {\bibfnamefont {J.}~\bibnamefont {Wu}}, \bibinfo {author}
  {\bibfnamefont {Q.}~\bibnamefont {Jiang}}, \bibinfo {author} {\bibfnamefont
  {M.}~\bibnamefont {Tang}}, \bibinfo {author} {\bibfnamefont {S.}~\bibnamefont
  {Shutts}}, \bibinfo {author} {\bibfnamefont {S.~N.}\ \bibnamefont {Elliott}},
  \bibinfo {author} {\bibfnamefont {A.}~\bibnamefont {Sobiesierski}}, \bibinfo
  {author} {\bibfnamefont {A.~J.}\ \bibnamefont {Seeds}}, \bibinfo {author}
  {\bibfnamefont {I.}~\bibnamefont {Ross}}, \bibinfo {author} {\bibfnamefont
  {P.~M.}\ \bibnamefont {Smowton}},\ and\ \bibinfo {author} {\bibfnamefont
  {H.}~\bibnamefont {Liu}},\ }\bibfield  {title} {\bibinfo {title}
  {Electrically pumped continuous-wave {III–V} quantum dot lasers on
  silicon},\ }\href@noop {} {\bibfield  {journal} {\bibinfo  {journal} {Nat.
  Photonics}\ }\textbf {\bibinfo {volume} {10}},\ \bibinfo {pages} {307}
  (\bibinfo {year} {2016})}\BibitemShut {NoStop}%
\bibitem [{\citenamefont {Vurgaftman}\ \emph {et~al.}(2001)\citenamefont
  {Vurgaftman}, \citenamefont {Meyer},\ and\ \citenamefont
  {Ram-Mohan}}]{Vurgaftman01p5815}%
  \BibitemOpen
  \bibfield  {author} {\bibinfo {author} {\bibfnamefont {I.}~\bibnamefont
  {Vurgaftman}}, \bibinfo {author} {\bibfnamefont {J.~R.}\ \bibnamefont
  {Meyer}},\ and\ \bibinfo {author} {\bibfnamefont {L.~R.}\ \bibnamefont
  {Ram-Mohan}},\ }\bibfield  {title} {\bibinfo {title} {Band parameters for
  {III–V} compound semiconductors and their alloys},\ }\href
  {https://doi.org/10.1063/1.1368156} {\bibfield  {journal} {\bibinfo
  {journal} {J. Appl. Phys.}\ }\textbf {\bibinfo {volume} {89}},\ \bibinfo
  {pages} {5815} (\bibinfo {year} {2001})}\BibitemShut {NoStop}%
\bibitem [{\citenamefont {Yang}\ \emph {et~al.}(2019)\citenamefont {Yang},
  \citenamefont {Li}, \citenamefont {Meng}, \citenamefont {Yu}, \citenamefont
  {Wang}, \citenamefont {Wang}, \citenamefont {Luo}, \citenamefont {Zhou},\
  and\ \citenamefont {Pan}}]{yang19p101305}%
  \BibitemOpen
  \bibfield  {author} {\bibinfo {author} {\bibfnamefont {W.}~\bibnamefont
  {Yang}}, \bibinfo {author} {\bibfnamefont {Y.}~\bibnamefont {Li}}, \bibinfo
  {author} {\bibfnamefont {F.}~\bibnamefont {Meng}}, \bibinfo {author}
  {\bibfnamefont {H.}~\bibnamefont {Yu}}, \bibinfo {author} {\bibfnamefont
  {M.}~\bibnamefont {Wang}}, \bibinfo {author} {\bibfnamefont {P.}~\bibnamefont
  {Wang}}, \bibinfo {author} {\bibfnamefont {G.}~\bibnamefont {Luo}}, \bibinfo
  {author} {\bibfnamefont {X.}~\bibnamefont {Zhou}},\ and\ \bibinfo {author}
  {\bibfnamefont {J.}~\bibnamefont {Pan}},\ }\bibfield  {title} {\bibinfo
  {title} {{III–V} compound materials and lasers on silicon},\ }\href
  {https://doi.org/10.1088/1674-4926/40/10/101305} {\bibfield  {journal}
  {\bibinfo  {journal} {J. Semicond.}\ }\textbf {\bibinfo {volume} {40}},\
  \bibinfo {pages} {101305} (\bibinfo {year} {2019})}\BibitemShut {NoStop}%
\bibitem [{III()}]{IIIV-on-Si}%
  \BibitemOpen
  \href@noop {} {\bibinfo {title} {\text{III-V} on \text{Silicon}}},\ \bibinfo
  {note}
  {\url{https://www.iqep.com/innovation/new-technology-platforms/iii-v-on-silicon/},
  (accessed on 2023--12)}\BibitemShut {NoStop}%
\bibitem [{\citenamefont {Junquera}\ and\ \citenamefont
  {Ghosez}(2003)}]{Junquera03p506}%
  \BibitemOpen
  \bibfield  {author} {\bibinfo {author} {\bibfnamefont {J.}~\bibnamefont
  {Junquera}}\ and\ \bibinfo {author} {\bibfnamefont {P.}~\bibnamefont
  {Ghosez}},\ }\bibfield  {title} {\bibinfo {title} {Critical thickness for
  ferroelectricity in perovskite ultrathin films},\ }\href
  {https://doi.org/10.1038/nature01501} {\bibfield  {journal} {\bibinfo
  {journal} {Nature}\ }\textbf {\bibinfo {volume} {422}},\ \bibinfo {pages}
  {506} (\bibinfo {year} {2003})}\BibitemShut {NoStop}%
\bibitem [{\citenamefont {Kresse}\ and\ \citenamefont
  {J}(1996{\natexlab{a}})}]{Kresse96p11169}%
  \BibitemOpen
  \bibfield  {author} {\bibinfo {author} {\bibfnamefont {G.}~\bibnamefont
  {Kresse}}\ and\ \bibinfo {author} {\bibfnamefont {F.}~\bibnamefont {J}},\
  }\bibfield  {title} {\bibinfo {title} {Efficient iterative schemes for ab
  initio total-energy calculations using a plane-wave basis set},\ }\href@noop
  {} {\bibfield  {journal} {\bibinfo  {journal} {Phys. Rev. B}\ }\textbf
  {\bibinfo {volume} {54}},\ \bibinfo {pages} {11169} (\bibinfo {year}
  {1996}{\natexlab{a}})}\BibitemShut {NoStop}%
\bibitem [{\citenamefont {Kresse}\ and\ \citenamefont
  {J}(1996{\natexlab{b}})}]{Kresse96p15}%
  \BibitemOpen
  \bibfield  {author} {\bibinfo {author} {\bibfnamefont {G.}~\bibnamefont
  {Kresse}}\ and\ \bibinfo {author} {\bibfnamefont {F.}~\bibnamefont {J}},\
  }\bibfield  {title} {\bibinfo {title} {Efficiency of ab-initio total energy
  calculations for metals and semiconductors using a plane-wave basis set},\
  }\href@noop {} {\bibfield  {journal} {\bibinfo  {journal} {Comput. Mater.
  Sci.}\ }\textbf {\bibinfo {volume} {6}},\ \bibinfo {pages} {15} (\bibinfo
  {year} {1996}{\natexlab{b}})}\BibitemShut {NoStop}%
\bibitem [{\citenamefont {Blochl}(1994)}]{Blochi94p17953}%
  \BibitemOpen
  \bibfield  {author} {\bibinfo {author} {\bibfnamefont {P.~E.}\ \bibnamefont
  {Blochl}},\ }\bibfield  {title} {\bibinfo {title} {Projector augmented-wave
  method},\ }\href@noop {} {\bibfield  {journal} {\bibinfo  {journal} {Phys.
  Rev. B}\ }\textbf {\bibinfo {volume} {50}},\ \bibinfo {pages} {17953}
  (\bibinfo {year} {1994})}\BibitemShut {NoStop}%
\bibitem [{\citenamefont {Perdew}\ \emph {et~al.}(2008)\citenamefont {Perdew},
  \citenamefont {Ruzsinszky}, \citenamefont {Csonka}, \citenamefont {Vydrov},
  \citenamefont {Scuseria}, \citenamefont {Constantin}, \citenamefont {Zhou},\
  and\ \citenamefont {Burke}}]{Perdew08p136406}%
  \BibitemOpen
  \bibfield  {author} {\bibinfo {author} {\bibfnamefont {J.~P.}\ \bibnamefont
  {Perdew}}, \bibinfo {author} {\bibfnamefont {A.}~\bibnamefont {Ruzsinszky}},
  \bibinfo {author} {\bibfnamefont {G.~I.}\ \bibnamefont {Csonka}}, \bibinfo
  {author} {\bibfnamefont {O.~A.}\ \bibnamefont {Vydrov}}, \bibinfo {author}
  {\bibfnamefont {G.~E.}\ \bibnamefont {Scuseria}}, \bibinfo {author}
  {\bibfnamefont {L.~A.}\ \bibnamefont {Constantin}}, \bibinfo {author}
  {\bibfnamefont {X.}~\bibnamefont {Zhou}},\ and\ \bibinfo {author}
  {\bibfnamefont {K.}~\bibnamefont {Burke}},\ }\bibfield  {title} {\bibinfo
  {title} {Restoring the density-gradient expansion for exchange in solids and
  surfaces},\ }\href {https://doi.org/10.1103/PhysRevLett.100.136406}
  {\bibfield  {journal} {\bibinfo  {journal} {Phys. Rev. Lett.}\ }\textbf
  {\bibinfo {volume} {100}},\ \bibinfo {pages} {136406} (\bibinfo {year}
  {2008})}\BibitemShut {NoStop}%
\bibitem [{\citenamefont {Nos{\'e}}(1984)}]{Nose84p511}%
  \BibitemOpen
  \bibfield  {author} {\bibinfo {author} {\bibfnamefont {S.}~\bibnamefont
  {Nos{\'e}}},\ }\bibfield  {title} {\bibinfo {title} {A unified formulation of
  the constant temperature molecular dynamics methods},\ }\href@noop {}
  {\bibfield  {journal} {\bibinfo  {journal} {J. Chem. Phys.}\ }\textbf
  {\bibinfo {volume} {81}},\ \bibinfo {pages} {511} (\bibinfo {year}
  {1984})}\BibitemShut {NoStop}%
\bibitem [{\citenamefont {Hoover}(1985)}]{Hoover85p1695}%
  \BibitemOpen
  \bibfield  {author} {\bibinfo {author} {\bibfnamefont {W.~G.}\ \bibnamefont
  {Hoover}},\ }\bibfield  {title} {\bibinfo {title} {Canonical dynamics:
  Equilibrium phase-space distributions},\ }\href@noop {} {\bibfield  {journal}
  {\bibinfo  {journal} {Phys. Rev. A}\ }\textbf {\bibinfo {volume} {31}},\
  \bibinfo {pages} {1695} (\bibinfo {year} {1985})}\BibitemShut {NoStop}%
\bibitem [{\citenamefont {Togo}\ and\ \citenamefont {Tanaka}(2015)}]{Togo15p1}%
  \BibitemOpen
  \bibfield  {author} {\bibinfo {author} {\bibfnamefont {A.}~\bibnamefont
  {Togo}}\ and\ \bibinfo {author} {\bibfnamefont {I.}~\bibnamefont {Tanaka}},\
  }\bibfield  {title} {\bibinfo {title} {First principles phonon calculations
  in materials science},\ }\href@noop {} {\bibfield  {journal} {\bibinfo
  {journal} {Scr. Mater.}\ }\textbf {\bibinfo {volume} {108}},\ \bibinfo
  {pages} {1} (\bibinfo {year} {2015})}\BibitemShut {NoStop}%
\bibitem [{\citenamefont {Wu}\ \emph {et~al.}(2011)\citenamefont {Wu},
  \citenamefont {Lagally},\ and\ \citenamefont {Liu}}]{Wu11p236101}%
  \BibitemOpen
  \bibfield  {author} {\bibinfo {author} {\bibfnamefont {D.}~\bibnamefont
  {Wu}}, \bibinfo {author} {\bibfnamefont {M.~G.}\ \bibnamefont {Lagally}},\
  and\ \bibinfo {author} {\bibfnamefont {F.}~\bibnamefont {Liu}},\ }\bibfield
  {title} {\bibinfo {title} {Stabilizing graphitic thin films of wurtzite
  materials by epitaxial strain},\ }\href
  {https://doi.org/10.1103/PhysRevLett.107.236101} {\bibfield  {journal}
  {\bibinfo  {journal} {Phys. Rev. Lett.}\ }\textbf {\bibinfo {volume} {107}},\
  \bibinfo {pages} {236101} (\bibinfo {year} {2011})}\BibitemShut {NoStop}%
\bibitem [{\citenamefont {\ifmmode~\mbox{\c{S}}\else \c{S}\fi{}ahin}\ \emph
  {et~al.}(2009)\citenamefont {\ifmmode~\mbox{\c{S}}\else \c{S}\fi{}ahin},
  \citenamefont {Cahangirov}, \citenamefont {Topsakal}, \citenamefont
  {Bekaroglu}, \citenamefont {Akturk}, \citenamefont {Senger},\ and\
  \citenamefont {Ciraci}}]{Sahin09p155453}%
  \BibitemOpen
  \bibfield  {author} {\bibinfo {author} {\bibfnamefont {H.}~\bibnamefont
  {\ifmmode~\mbox{\c{S}}\else \c{S}\fi{}ahin}}, \bibinfo {author}
  {\bibfnamefont {S.}~\bibnamefont {Cahangirov}}, \bibinfo {author}
  {\bibfnamefont {M.}~\bibnamefont {Topsakal}}, \bibinfo {author}
  {\bibfnamefont {E.}~\bibnamefont {Bekaroglu}}, \bibinfo {author}
  {\bibfnamefont {E.}~\bibnamefont {Akturk}}, \bibinfo {author} {\bibfnamefont
  {R.~T.}\ \bibnamefont {Senger}},\ and\ \bibinfo {author} {\bibfnamefont
  {S.}~\bibnamefont {Ciraci}},\ }\bibfield  {title} {\bibinfo {title}
  {Monolayer honeycomb structures of group-{IV} elements and {III-V} binary
  compounds: First-principles calculations},\ }\href
  {https://doi.org/10.1103/PhysRevB.80.155453} {\bibfield  {journal} {\bibinfo
  {journal} {Phys. Rev. B}\ }\textbf {\bibinfo {volume} {80}},\ \bibinfo
  {pages} {155453} (\bibinfo {year} {2009})}\BibitemShut {NoStop}%
\bibitem [{\citenamefont {Zhuang}\ \emph {et~al.}(2013)\citenamefont {Zhuang},
  \citenamefont {Singh},\ and\ \citenamefont {Hennig}}]{Zhuang13p165415}%
  \BibitemOpen
  \bibfield  {author} {\bibinfo {author} {\bibfnamefont {H.~L.}\ \bibnamefont
  {Zhuang}}, \bibinfo {author} {\bibfnamefont {A.~K.}\ \bibnamefont {Singh}},\
  and\ \bibinfo {author} {\bibfnamefont {R.~G.}\ \bibnamefont {Hennig}},\
  }\bibfield  {title} {\bibinfo {title} {Computational discovery of
  single-layer {III-V} materials},\ }\href
  {https://doi.org/10.1103/PhysRevB.87.165415} {\bibfield  {journal} {\bibinfo
  {journal} {Phys. Rev. B}\ }\textbf {\bibinfo {volume} {87}},\ \bibinfo
  {pages} {165415} (\bibinfo {year} {2013})}\BibitemShut {NoStop}%
\bibitem [{\citenamefont {Gao}\ and\ \citenamefont
  {Gao}(2017)}]{Gao17p1600412}%
  \BibitemOpen
  \bibfield  {author} {\bibinfo {author} {\bibfnamefont {R.}~\bibnamefont
  {Gao}}\ and\ \bibinfo {author} {\bibfnamefont {Y.}~\bibnamefont {Gao}},\
  }\bibfield  {title} {\bibinfo {title} {Piezoelectricity in two-dimensional
  group {III}-{V} buckled honeycomb monolayers},\ }\href
  {https://doi.org/10.1002/pssr.201600412} {\bibfield  {journal} {\bibinfo
  {journal} {Phys. Status Solidi {RRL}}\ }\textbf {\bibinfo {volume} {11}},\
  \bibinfo {pages} {1600412} (\bibinfo {year} {2017})}\BibitemShut {NoStop}%
\bibitem [{\citenamefont {Shirane}(1952)}]{Shirane52p219}%
  \BibitemOpen
  \bibfield  {author} {\bibinfo {author} {\bibfnamefont {G.}~\bibnamefont
  {Shirane}},\ }\bibfield  {title} {\bibinfo {title} {Ferroelectricity and
  antiferroelectricity in ceramic {PbZrO$_3$} containing {Ba} or {Sr}},\ }\href
  {https://doi.org/10.1103/PhysRev.86.219} {\bibfield  {journal} {\bibinfo
  {journal} {Phys. Rev.}\ }\textbf {\bibinfo {volume} {86}},\ \bibinfo {pages}
  {219} (\bibinfo {year} {1952})}\BibitemShut {NoStop}%
\bibitem [{\citenamefont {Huang}\ \emph
  {et~al.}(2022{\natexlab{a}})\citenamefont {Huang}, \citenamefont {Hu},\ and\
  \citenamefont {Liu}}]{Huang22p144106}%
  \BibitemOpen
  \bibfield  {author} {\bibinfo {author} {\bibfnamefont {J.}~\bibnamefont
  {Huang}}, \bibinfo {author} {\bibfnamefont {Y.}~\bibnamefont {Hu}},\ and\
  \bibinfo {author} {\bibfnamefont {S.}~\bibnamefont {Liu}},\ }\bibfield
  {title} {\bibinfo {title} {Origin of ferroelectricity in magnesium-doped zinc
  oxide},\ }\href {https://doi.org/10.1103/PhysRevB.106.144106} {\bibfield
  {journal} {\bibinfo  {journal} {Phys. Rev. B}\ }\textbf {\bibinfo {volume}
  {106}},\ \bibinfo {pages} {144106} (\bibinfo {year}
  {2022}{\natexlab{a}})}\BibitemShut {NoStop}%
\bibitem [{\citenamefont {Oganov}\ and\ \citenamefont
  {Glass}(2006)}]{Oganov06p244704}%
  \BibitemOpen
  \bibfield  {author} {\bibinfo {author} {\bibfnamefont {A.~R.}\ \bibnamefont
  {Oganov}}\ and\ \bibinfo {author} {\bibfnamefont {C.~W.}\ \bibnamefont
  {Glass}},\ }\bibfield  {title} {\bibinfo {title} {Crystal structure
  prediction using ab initio evolutionary techniques: Principles and
  applications},\ }\href {https://doi.org/10.1063/1.2210932} {\bibfield
  {journal} {\bibinfo  {journal} {J. Chem. Phys.}\ }\textbf {\bibinfo {volume}
  {124}},\ \bibinfo {pages} {244704} (\bibinfo {year} {2006})}\BibitemShut
  {NoStop}%
\bibitem [{\citenamefont {Lyakhov}\ \emph {et~al.}(2013)\citenamefont
  {Lyakhov}, \citenamefont {Oganov}, \citenamefont {Stokes},\ and\
  \citenamefont {Zhu}}]{Lyakhov13p1172}%
  \BibitemOpen
  \bibfield  {author} {\bibinfo {author} {\bibfnamefont {A.~O.}\ \bibnamefont
  {Lyakhov}}, \bibinfo {author} {\bibfnamefont {A.~R.}\ \bibnamefont {Oganov}},
  \bibinfo {author} {\bibfnamefont {H.~T.}\ \bibnamefont {Stokes}},\ and\
  \bibinfo {author} {\bibfnamefont {Q.}~\bibnamefont {Zhu}},\ }\bibfield
  {title} {\bibinfo {title} {New developments in evolutionary structure
  prediction algorithm {USPEX}},\ }\href
  {https://doi.org/10.1016/j.cpc.2012.12.009} {\bibfield  {journal} {\bibinfo
  {journal} {Comput. Phys. Commun.}\ }\textbf {\bibinfo {volume} {184}},\
  \bibinfo {pages} {1172} (\bibinfo {year} {2013})}\BibitemShut {NoStop}%
\bibitem [{\citenamefont {Oganov}\ \emph {et~al.}(2011)\citenamefont {Oganov},
  \citenamefont {Lyakhov},\ and\ \citenamefont {Valle}}]{Oganov11p227}%
  \BibitemOpen
  \bibfield  {author} {\bibinfo {author} {\bibfnamefont {A.~R.}\ \bibnamefont
  {Oganov}}, \bibinfo {author} {\bibfnamefont {A.~O.}\ \bibnamefont
  {Lyakhov}},\ and\ \bibinfo {author} {\bibfnamefont {M.}~\bibnamefont
  {Valle}},\ }\bibfield  {title} {\bibinfo {title} {How evolutionary crystal
  structure prediction works{\textemdash}and why},\ }\href
  {https://doi.org/10.1021/ar1001318} {\bibfield  {journal} {\bibinfo
  {journal} {Acc. Chem. Res.}\ }\textbf {\bibinfo {volume} {44}},\ \bibinfo
  {pages} {227} (\bibinfo {year} {2011})}\BibitemShut {NoStop}%
\bibitem [{\citenamefont {Adhikari}\ and\ \citenamefont
  {Fu}(2019)}]{Adhikari19p104101}%
  \BibitemOpen
  \bibfield  {author} {\bibinfo {author} {\bibfnamefont {R.}~\bibnamefont
  {Adhikari}}\ and\ \bibinfo {author} {\bibfnamefont {H.}~\bibnamefont {Fu}},\
  }\bibfield  {title} {\bibinfo {title} {Hyperferroelectricity in zno: Evidence
  from analytic formulation and numerical calculations},\ }\href
  {https://doi.org/10.1103/PhysRevB.99.104101} {\bibfield  {journal} {\bibinfo
  {journal} {Phys. Rev. B}\ }\textbf {\bibinfo {volume} {99}},\ \bibinfo
  {pages} {104101} (\bibinfo {year} {2019})}\BibitemShut {NoStop}%
\bibitem [{\citenamefont {Qiu}\ \emph {et~al.}(2021)\citenamefont {Qiu},
  \citenamefont {Ma}, \citenamefont {Liu},\ and\ \citenamefont
  {Fu}}]{Qiu21p064112}%
  \BibitemOpen
  \bibfield  {author} {\bibinfo {author} {\bibfnamefont {S.}~\bibnamefont
  {Qiu}}, \bibinfo {author} {\bibfnamefont {L.}~\bibnamefont {Ma}}, \bibinfo
  {author} {\bibfnamefont {S.}~\bibnamefont {Liu}},\ and\ \bibinfo {author}
  {\bibfnamefont {H.}~\bibnamefont {Fu}},\ }\bibfield  {title} {\bibinfo
  {title} {Possible existence of tristable polarization states in
  ${\mathrm{linbo}}_{3}$ under an open-circuit boundary condition},\ }\href
  {https://doi.org/10.1103/PhysRevB.104.064112} {\bibfield  {journal} {\bibinfo
   {journal} {Phys. Rev. B}\ }\textbf {\bibinfo {volume} {104}},\ \bibinfo
  {pages} {064112} (\bibinfo {year} {2021})}\BibitemShut {NoStop}%
\bibitem [{\citenamefont {Garrity}\ \emph {et~al.}(2014)\citenamefont
  {Garrity}, \citenamefont {Rabe},\ and\ \citenamefont
  {Vanderbilt}}]{Garrity14p127601}%
  \BibitemOpen
  \bibfield  {author} {\bibinfo {author} {\bibfnamefont {K.~F.}\ \bibnamefont
  {Garrity}}, \bibinfo {author} {\bibfnamefont {K.~M.}\ \bibnamefont {Rabe}},\
  and\ \bibinfo {author} {\bibfnamefont {D.}~\bibnamefont {Vanderbilt}},\
  }\bibfield  {title} {\bibinfo {title} {Hyperferroelectrics: Proper
  ferroelectrics with persistent polarization},\ }\href
  {https://doi.org/10.1103/PhysRevLett.112.127601} {\bibfield  {journal}
  {\bibinfo  {journal} {Phys. Rev. Lett.}\ }\textbf {\bibinfo {volume} {112}},\
  \bibinfo {pages} {127601} (\bibinfo {year} {2014})}\BibitemShut {NoStop}%
\bibitem [{\citenamefont {Agapito}\ \emph {et~al.}(2013)\citenamefont
  {Agapito}, \citenamefont {Ferretti}, \citenamefont {Calzolari}, \citenamefont
  {Curtarolo},\ and\ \citenamefont {Nardelli}}]{Agapito13p165127}%
  \BibitemOpen
  \bibfield  {author} {\bibinfo {author} {\bibfnamefont {L.~A.}\ \bibnamefont
  {Agapito}}, \bibinfo {author} {\bibfnamefont {A.}~\bibnamefont {Ferretti}},
  \bibinfo {author} {\bibfnamefont {A.}~\bibnamefont {Calzolari}}, \bibinfo
  {author} {\bibfnamefont {S.}~\bibnamefont {Curtarolo}},\ and\ \bibinfo
  {author} {\bibfnamefont {M.~B.}\ \bibnamefont {Nardelli}},\ }\bibfield
  {title} {\bibinfo {title} {Effective and accurate representation of extended
  {Bloch} states on finite {Hilbert} spaces},\ }\href@noop {} {\bibfield
  {journal} {\bibinfo  {journal} {Phys. Rev. B}\ }\textbf {\bibinfo {volume}
  {88}},\ \bibinfo {pages} {165127} (\bibinfo {year} {2013})}\BibitemShut
  {NoStop}%
\bibitem [{\citenamefont {Lee}\ and\ \citenamefont {Son}(2020)}]{Lee20p043410}%
  \BibitemOpen
  \bibfield  {author} {\bibinfo {author} {\bibfnamefont {S.-H.}\ \bibnamefont
  {Lee}}\ and\ \bibinfo {author} {\bibfnamefont {Y.-W.}\ \bibnamefont {Son}},\
  }\bibfield  {title} {\bibinfo {title} {First-principles approach with a
  pseudohybrid density functional for extended {Hubbard} interactions},\ }\href
  {https://doi.org/10.1103/PhysRevResearch.2.043410} {\bibfield  {journal}
  {\bibinfo  {journal} {Phys. Rev. Research}\ }\textbf {\bibinfo {volume}
  {2}},\ \bibinfo {pages} {043410} (\bibinfo {year} {2020})}\BibitemShut
  {NoStop}%
\bibitem [{\citenamefont {Tancogne-Dejean}\ and\ \citenamefont
  {Rubio}(2020)}]{Tancogne-Dejean20p155117}%
  \BibitemOpen
  \bibfield  {author} {\bibinfo {author} {\bibfnamefont {N.}~\bibnamefont
  {Tancogne-Dejean}}\ and\ \bibinfo {author} {\bibfnamefont {A.}~\bibnamefont
  {Rubio}},\ }\bibfield  {title} {\bibinfo {title} {Parameter-free hybridlike
  functional based on an extended {Hubbard} model: {{DFT}+$U$+$V$}},\ }\href
  {https://doi.org/10.1103/PhysRevB.102.155117} {\bibfield  {journal} {\bibinfo
   {journal} {Phys. Rev. B}\ }\textbf {\bibinfo {volume} {102}},\ \bibinfo
  {pages} {155117} (\bibinfo {year} {2020})}\BibitemShut {NoStop}%
\bibitem [{\citenamefont {Krukau}\ \emph {et~al.}(2006)\citenamefont {Krukau},
  \citenamefont {Vydrov}, \citenamefont {Izmaylov},\ and\ \citenamefont
  {Scuseria}}]{Krukau06p224106}%
  \BibitemOpen
  \bibfield  {author} {\bibinfo {author} {\bibfnamefont {A.~V.}\ \bibnamefont
  {Krukau}}, \bibinfo {author} {\bibfnamefont {O.~A.}\ \bibnamefont {Vydrov}},
  \bibinfo {author} {\bibfnamefont {A.~F.}\ \bibnamefont {Izmaylov}},\ and\
  \bibinfo {author} {\bibfnamefont {G.~E.}\ \bibnamefont {Scuseria}},\
  }\bibfield  {title} {\bibinfo {title} {Influence of the exchange screening
  parameter on the performance of screened hybrid functionals},\ }\href
  {https://doi.org/10.1063/1.2404663} {\bibfield  {journal} {\bibinfo
  {journal} {J. Chem. Phys.}\ }\textbf {\bibinfo {volume} {125}},\ \bibinfo
  {pages} {224106} (\bibinfo {year} {2006})}\BibitemShut {NoStop}%
\bibitem [{\citenamefont {Huang}\ \emph {et~al.}(2020)\citenamefont {Huang},
  \citenamefont {Lee}, \citenamefont {Son}, \citenamefont {Supka},\ and\
  \citenamefont {Liu}}]{Huang20p165157}%
  \BibitemOpen
  \bibfield  {author} {\bibinfo {author} {\bibfnamefont {J.}~\bibnamefont
  {Huang}}, \bibinfo {author} {\bibfnamefont {S.-H.}\ \bibnamefont {Lee}},
  \bibinfo {author} {\bibfnamefont {Y.-W.}\ \bibnamefont {Son}}, \bibinfo
  {author} {\bibfnamefont {A.}~\bibnamefont {Supka}},\ and\ \bibinfo {author}
  {\bibfnamefont {S.}~\bibnamefont {Liu}},\ }\bibfield  {title} {\bibinfo
  {title} {First-principles study of two-dimensional ferroelectrics using
  self-consistent hubbard parameters},\ }\href
  {https://doi.org/10.1103/PhysRevB.102.165157} {\bibfield  {journal} {\bibinfo
   {journal} {Phys. Rev. B}\ }\textbf {\bibinfo {volume} {102}},\ \bibinfo
  {pages} {165157} (\bibinfo {year} {2020})}\BibitemShut {NoStop}%
\bibitem [{\citenamefont {Ke}\ \emph {et~al.}(2021)\citenamefont {Ke},
  \citenamefont {Huang},\ and\ \citenamefont {Liu}}]{Ke21p3387}%
  \BibitemOpen
  \bibfield  {author} {\bibinfo {author} {\bibfnamefont {C.}~\bibnamefont
  {Ke}}, \bibinfo {author} {\bibfnamefont {J.}~\bibnamefont {Huang}},\ and\
  \bibinfo {author} {\bibfnamefont {S.}~\bibnamefont {Liu}},\ }\bibfield
  {title} {\bibinfo {title} {Two-dimensional ferroelectric metal for
  electrocatalysis},\ }\href {https://doi.org/10.1039/d1mh01556g} {\bibfield
  {journal} {\bibinfo  {journal} {Mater. Horiz.}\ }\textbf {\bibinfo {volume}
  {8}},\ \bibinfo {pages} {3387} (\bibinfo {year} {2021})}\BibitemShut
  {NoStop}%
\bibitem [{\citenamefont {Huang}\ \emph
  {et~al.}(2022{\natexlab{b}})\citenamefont {Huang}, \citenamefont {Duan},
  \citenamefont {Jeon}, \citenamefont {Kim}, \citenamefont {Zhou},
  \citenamefont {Li},\ and\ \citenamefont {Liu}}]{Huang22p1440}%
  \BibitemOpen
  \bibfield  {author} {\bibinfo {author} {\bibfnamefont {J.}~\bibnamefont
  {Huang}}, \bibinfo {author} {\bibfnamefont {X.}~\bibnamefont {Duan}},
  \bibinfo {author} {\bibfnamefont {S.}~\bibnamefont {Jeon}}, \bibinfo {author}
  {\bibfnamefont {Y.}~\bibnamefont {Kim}}, \bibinfo {author} {\bibfnamefont
  {J.}~\bibnamefont {Zhou}}, \bibinfo {author} {\bibfnamefont {J.}~\bibnamefont
  {Li}},\ and\ \bibinfo {author} {\bibfnamefont {S.}~\bibnamefont {Liu}},\
  }\bibfield  {title} {\bibinfo {title} {On-demand quantum spin hall insulators
  controlled by two-dimensional ferroelectricity},\ }\href
  {https://doi.org/10.1039/d2mh00334a} {\bibfield  {journal} {\bibinfo
  {journal} {Mater. Horiz.}\ }\textbf {\bibinfo {volume} {9}},\ \bibinfo
  {pages} {1440} (\bibinfo {year} {2022}{\natexlab{b}})}\BibitemShut {NoStop}%
\bibitem [{\citenamefont {Duan}\ \emph {et~al.}(2021)\citenamefont {Duan},
  \citenamefont {Huang}, \citenamefont {Xu},\ and\ \citenamefont
  {Liu}}]{Duan21p2316}%
  \BibitemOpen
  \bibfield  {author} {\bibinfo {author} {\bibfnamefont {X.}~\bibnamefont
  {Duan}}, \bibinfo {author} {\bibfnamefont {J.}~\bibnamefont {Huang}},
  \bibinfo {author} {\bibfnamefont {B.}~\bibnamefont {Xu}},\ and\ \bibinfo
  {author} {\bibfnamefont {S.}~\bibnamefont {Liu}},\ }\bibfield  {title}
  {\bibinfo {title} {A two-dimensional multiferroic metal with voltage-tunable
  magnetization and metallicity},\ }\href {https://doi.org/10.1039/d1mh00939g}
  {\bibfield  {journal} {\bibinfo  {journal} {Mater. Horiz}\ }\textbf {\bibinfo
  {volume} {8}},\ \bibinfo {pages} {2316} (\bibinfo {year} {2021})}\BibitemShut
  {NoStop}%
\end{thebibliography}%
\clearpage
\newpage

\end{document}